%% file: main.tex
\documentclass[sigconf]{acmart}

\AtBeginDocument{%
  \providecommand\BibTeX{{%
    \normalfont B\kern-0.5em{\scshape i\kern-0.25em b}\kern-0.8em\TeX}}}

\copyrightyear{2020}
\acmYear{2020}
\setcopyright{acmcopyright}
\acmConference[ICSE '20]{42nd International Conference on Software Engineering}{May 23--29, 2020}{Seoul, Republic of Korea}
\acmBooktitle{42nd International Conference on Software Engineering (ICSE '20), May 23--29, 2020, Seoul, Republic of Korea}
\acmPrice{15.00}
\acmDOI{10.1145/3377811.3380400}
\acmISBN{978-1-4503-7121-6/20/05}

\include{includes}

\include{macros}

\include{pygments}

\lstdefinestyle{example} {
  frame=tb,
  basicstyle=\footnotesize\ttfamily,
  numbers=left,
  language=C
}

\usepackage{flushend}
\begin{document}

\title{Testing DNN Image Classifiers for Confusion \& Bias Errors}

 \author{Yuchi Tian}
 \authornote{Both are first authors, and contributed equally to this research.}
 \affiliation{%
  \institution{Columbia University}}
\email{yuchi.tian@columbia.edu}

 \author{Ziyuan Zhong}\authornotemark[1]
 \affiliation{%
  \institution{Columbia University}}
 \email{ziyuan.zhong@columbia.edu} 
  
   \author{Vicente Ordonez}
 \affiliation{%
  \institution{University of Virginia}}
  \email{vicente@virginia.edu}
  
   \author{Gail Kaiser}
 \affiliation{%
  \institution{Columbia University}}
  \email{kaiser@cs.columbia.edu}
  
   \author{Baishakhi Ray}
 \affiliation{%
  \institution{Columbia University}}
  \email{rayb@cs.columbia.edu}

\begin{abstract}
Image classifiers are an important component of today's software, from consumer and business applications to safety-critical domains. The advent of Deep Neural Networks (DNNs) is the key catalyst behind such wide-spread success. However, wide adoption comes with serious concerns about the robustness of software systems dependent on DNNs for image classification, as several severe erroneous behaviors have been reported under sensitive and critical circumstances. We argue that developers need to rigorously test their software's image classifiers and delay deployment until acceptable. We present an approach to testing image classifier robustness based on class property violations.

We found that many of the reported erroneous cases in popular DNN image classifiers occur because the trained models confuse one class with another or show biases towards some classes over others. These bugs usually violate some class properties of one or more of those classes. Most DNN testing techniques focus on per-image violations, so fail to detect class-level confusions or biases.  

We developed a testing technique to automatically detect  class-based confusion and bias errors in DNN-driven image classification software.  We evaluated our implementation, DeepInspect, on several popular image classifiers with precision up to 100\% (avg.~72.6\%) for confusion errors, and up to 84.3\% (avg.~66.8\%) for bias errors. DeepInspect found hundreds of classification mistakes in widely-used models, many exposing errors indicating confusion or bias. 
\end{abstract}

\ccsdesc[500]{Software and its engineering~Software testing and debugging}
\ccsdesc[300]{Computing methodologies~Neural networks}

\keywords{whitebox testing, deep learning, DNNs, image classifiers, bias}

\maketitle

\input{body.tex}


\begin{acks}
This work is supported in part by NSF CNS-1563555, CCF-1815494, CNS-1842456, CCF-1845893, and CCF-1822965.
Any opinions, findings, conclusions, or recommendations expressed herein are those
of the authors, and do not necessarily reflect those of the US Government or NSF. 
\end{acks}

\balance


\bibliographystyle{ACM-Reference-Format}
\bibliography{main}










\end{document}

%% file: includes.tex
\usepackage{amsmath}

\PassOptionsToPackage{hyphens}{url}\usepackage{hyperref}
\usepackage[noabbrev]{cleveref}
\hypersetup{colorlinks,urlcolor=blue,citecolor=black}
\usepackage[T1]{fontenc}
\usepackage[export]{adjustbox}
\usepackage[flushleft]{threeparttable}
\usepackage[linesnumbered, ruled]{algorithm2e}
\usepackage[turnon]{notes}
\usepackage[utf8]{inputenc}
\usepackage{algorithmic}

\DeclareMathOperator{\typeaconf}{type1conf}
\DeclareMathOperator{\typebconf}{type2conf}
\DeclareMathOperator{\mean}{mean}
\DeclareMathOperator{\prob}{P}

\usepackage{amsfonts}
\usepackage[export]{adjustbox}

\usepackage{array}
\usepackage{balance}
\usepackage{bm}
\usepackage{blkarray}
\usepackage{booktabs}
\usepackage{caption}
\usepackage{color}
\usepackage{xcolor,colortbl}
\usepackage{comment}

\usepackage{enumitem}
\usepackage{eqparbox}
\usepackage{fancybox}
\usepackage{fancyvrb}
\usepackage{framed}
\usepackage{graphicx}
\usepackage{ifthen}
\usepackage{listings}
\usepackage{listofitems}
\usepackage{makecell}
\usepackage{mdwmath}
\usepackage{mdwtab}
\usepackage{microtype}
\usepackage{multirow}
\usepackage{ragged2e}
\usepackage{stackengine}
\usepackage{stfloats}
\usepackage{subcaption} 
\usepackage{url}
\usepackage{wrapfig}
\usepackage{xspace}
\usepackage{nicefrac}

\usepackage{moreverb}

\everymath{\displaystyle}

\newcommand{\rom}[1]{\uppercase\expandafter{\romannumeral #1\relax}}

\newcommand{\etal}{\hbox{\emph{et al.}}\xspace}
\newcommand{\eg}{\hbox{\emph{e.g.,}}\xspace}
\newcommand{\ie}{\hbox{\emph{i.e.}}\xspace}

\newcommand{\wrt}{\hbox{\emph{w.r.t.}}\xspace}

\newcommand{\etc}{\hbox{\emph{etc.}}\xspace}

\definecolor{gray50}{gray}{.5}
\definecolor{gray40}{gray}{.6}
\definecolor{gray30}{gray}{.7}
\definecolor{gray20}{gray}{.8}
\definecolor{gray10}{gray}{.9}
\definecolor{gray05}{gray}{.95}

\newlength\Linewidth
\def\findlength{\setlength\Linewidth\linewidth
\addtolength\Linewidth{-4\fboxrule}
\addtolength\Linewidth{-3\fboxsep}
}
\newenvironment{examplebox}{\par\begingroup
   \setlength{\fboxsep}{5pt}\findlength
   \setbox0=\vbox\bgroup\noindent
   \hsize=0.95\linewidth
   \begin{minipage}{0.95\linewidth}\normalsize}
    {\end{minipage}\egroup
    \textcolor{gray20}{\fboxsep1.5pt\fbox
     {\fboxsep5pt\colorbox{gray05}{\normalcolor\box0}}}
    \endgroup\par\noindent
    \normalcolor\ignorespacesafterend}

\newcounter{RQCounter}
\newcounter{RQACounter}

\newcommand{\RQ}[2]{%
\refstepcounter{RQCounter} \label{#1}
 \noindent
 \textbf{RQ\arabic{RQCounter}.~#2}
}

\newcommand{\RS}[2]{%
\begin{framed}%
\filbreak
\textbf{Result {\ref{#1}}:~}{\emph {#2}}%
\end{framed}
}

\definecolor{javared}{rgb}{0.6,0,0} 
\definecolor{javagreen}{rgb}{0.25,0.5,0.35} 
\definecolor{javapurple}{rgb}{0.5,0,0.35} 
\definecolor{javadocblue}{rgb}{0.25,0.35,0.75} 

\lstdefinestyle{customc}{
  belowcaptionskip=\baselineskip,
  breaklines=true,
  xleftmargin=\parindent,
  language=java,
  showstringspaces=false,
  basicstyle=\scriptsize\ttfamily,
  keywordstyle=\bfseries\color{javapurple},
  commentstyle=\itshape\blue,
}

\DeclareCaptionLabelFormat{andimage}{#1~#2  \&  \figurename~\thefigure}

%% file: macros.tex
\newcommand{\Comment}[1]{}

\newcommand{\tool}{DeepInspect\xspace}
\newcommand{\ta}{Type1 confusions\xspace}
\newcommand{\tb}{Type2 confusions\xspace}
\newcommand{\navd}{\textsc{napvd}\xspace}

\newcommand{\napm}{Neuron Activation Probability Matrix\xspace}

\newcommand{\coco}{COCO~}
\newcommand{\cifar}{CIFAR-100~}

\newcommand{\ddata}{experimental data\xspace}

\newcommand{\ab}{\textrm{avg\_bias}\xspace}
\newcommand{\cd}{\textrm{cd}\xspace}
\newcommand{\acd}{\textrm{avg\_cd}\xspace}

\newcommand{\rqa}{Can \tool distinguish between different classes?~}
\newcommand{\rqb}{Can \tool identify the confusion errors?~}
\newcommand{\rqc}{Can \tool identify the bias errors?~}

\makeatother

%% file: pygments.tex
\newcommand\red[1]{\textcolor[rgb]{1.00,0.00,0.00}{#1}}

\newcommand\blue[1]{\textcolor[rgb]{0.00,0.00,1.00}{{#1}}}

%% file: body.tex

\section{Introduction}
\label{sec:intro}

Image classification has a plethora of applications in software for safety-critical domains such as self-driving cars, medical diagnosis, \etc Even day-to-day consumer software includes image classifiers, such as Google Photo search and Facebook image tagging. Image classification is a well-studied problem in computer vision, where a model is trained to classify an image into single or multiple predefined categories~\cite{kamavisdar2013survey}. 
Deep Neural Networks (DNNs) have enabled major breakthroughs in image classification tasks over the past few years, 
sometimes even matching human-level accuracy under some conditions~\cite{he2016deep}, which has led to their ubiquity in modern software.

However, in spite of such spectacular success, DNN-based image classification models, like traditional software, are known to have serious bugs.  For example, Google faced backlash in 2015 due to a notorious error in its photo-tagging app, which tagged pictures of dark-skinned people as ``gorillas''~\cite{google_tag}. Analogous to traditional software bugs, the Software Engineering (SE) literature denotes these classification errors as {\em model bugs}~\cite{Ma:2018:MAN:3236024.3236082}, which can arise due to either imperfect model structure or inadequate training data.

At a high-level, these bugs can affect either an {\em individual image}, where a particular image is mis-classified (\eg a particular skier is mistaken as a part of a mountain), 
or an {\em image class}, where a class of images is more likely to be mis-classified (\eg dark-skinned people are more likely to be misclassified), as shown in Table~\ref{tab:example}. The latter bugs are specific to a whole {\em class} of images rather than individual images, implying systematic bugs rather than the DNN equivalent of off-by-one errors. While much effort from the SE literature on Neural Network testing has focused on identifying individual-level violations\textemdash using white-box~\cite{pei2017deepxplore,ma2018deepgauge, kim2019guiding, wang2019adversarial}, grey-box~\cite{tian2017deeptest,Ma:2018:MAN:3236024.3236082}, or concolic testing~\cite{sun2018concolic}, detection of class-level violations remains relatively less explored. This paper focuses on automatically detecting such class-level bugs, so they can be fixed. 

\input{tables/example.tex}

After manual investigation of some public reports describing the class-level violations listed in Table~\ref{tab:example}, we determined two root causes: (i) \textbf{Confusion}: The model cannot differentiate one class from another. For example, Google Photos confuses skier and mountain~\cite{google_photo}.  (ii) \textbf{Bias}: The model shows disparate outcomes between two related groups. For example, Zhao \etal in their paper ``Men also like  shopping''~\cite{zhao2017men}, find classification bias in favor of women on activities like shopping, cooking, washing, \etc We further notice that some class-level properties are violated in both kinds of cases. For example, in the case of {\em confusion errors}, the classification error-rate between the objects of two classes, say, skier and mountain, is significantly higher than the overall classification error rate of the model. Similarly, in the bias scenario reported by Zhao \etal, a DNN model should not have different error rates while classifying the gender of a person in the shopping category. Unlike individual image properties, this is a class property affecting all the shopping images with men or women. Any violation of such a property by definition affects the whole class although not necessarily every image in that class, \eg a man is more prone to be predicted as a woman when he is shopping, even though some individual images of a man shopping may still be predicted correctly.  Thus, we need a class-level approach to testing image classifier software for confusion and bias errors.

The bugs in a DNN model occur due to sub-optimal interactions between the model structure and the training data~\cite{Ma:2018:MAN:3236024.3236082}. 
To capture such interactions, the literature has proposed various metrics primarily based on either neuron activations~\cite{pei2017deepxplore, ma2018deepgauge, kim2019guiding} or feature vectors~\cite{Ma:2018:MAN:3236024.3236082, pmlr-v97-odena19a}. 
However, these techniques are primarily targeted at the individual image level. To detect class-level violations, we abstract away such model-data interactions at the class level and analyze the inter-class interactions using that new abstraction. To this end, we propose a metric using neuron activations and a baseline metric using weight vectors of the feature embedding to capture the class abstraction.

For a set of test input images, we compute the probability of activation of a neuron per predicted class.
Thus, for each class, we create a vector of neuron activations where each vector element corresponds to a neuron activation probability. If the distance between the two vectors for two different classes is too close, compared to other class-vector pairs, that means the DNN under test may not effectively distinguish between those two classes. 
Motivated by MODE’s technique~\cite{Ma:2018:MAN:3236024.3236082}, we further create a baseline where each class is represented by the corresponding weight vector of the last linear layer of the model under test.

We evaluate our methodology for both single- and multi-label classification models in eight different settings. Our experiments demonstrate that \tool can efficiently detect both Bias and Confusion errors in popular neural image classifiers.  
 We further check whether \tool can detect such classification errors in state-of-the-art models designed to be robust against norm-bounded adversarial attacks~\cite{NIPS2018_8060}; \tool finds hundreds of errors proving the need for orthogonal testing strategies to detect such class-level mispredictions. 
Unlike some other DNN testing techniques~\cite{tian2017deeptest,pei2017deepxplore,pmlr-v97-odena19a,sun2018concolic}, \tool does not need to generate additional transformed (synthetic) images to find these errors.   
The primary contributions of this paper are: 

\begin{itemize}[leftmargin=*]
    \item 
    We propose a novel neuron-coverage metric to automatically detect class-level violations (confusion and bias errors) in DNN-based visual recognition models for image classification. 
    \item 
    We implemented our metric and underlying techniques in \tool.
    \item
    We evaluated \tool and found many errors in widely-used DNN models with precision up to 100\% (avg.~72.6\%) for confusion errors and up to 84.3\% (avg.~66.8\%) for bias errors.
\end{itemize}

Our code is available at \url{https://github.com/ARiSE-Lab/DeepInspect}.
The errors reported by \tool are available at: \url{https://www.ariselab.info/deepinspect}.

\section{DNN Background}
\label{sec:background}

Deep Neural Networks (DNNs) are a popular type of machine learning model loosely inspired by the neural networks of human brains. A DNN model learns the logic to perform a software task from a large set of \emph{training examples}. For example, an image recognition model learns to recognize {\tt cows} through being shown (trained with) many sample images of cows.

A typical "feed-forward" DNN consists of a set of connected computational units, referred as {\em neurons}, that are arranged sequentially in a series of {\em layers}.  
The neurons in sequential layers are connected to each other through \textit{edges}.
Each edge has a corresponding weight. 
Each neuron applies $\sigma$, a \textit{nonlinear activation function} (\eg ReLU~\cite{nair2010rectified}, Sigmoid~\cite{Mitchell:1997:ML:541177}), to the incoming values on its input edges and sends the results on its output edges to the next layer of connected neurons. 
For image classification, convolutional neural networks (CNNs)~\cite{lecun1998gradient}, a specific type of DNN,  are typically used. CNNs consist of layers with local spatial connectivity and sets of neurons with shared parameters.

When implementing a DNN application, developers 
typically start with a set of annotated \ddata and divide it into three sets: 
(i) training: to construct the DNN model in a supervised setting, meaning the training data is labeled (\eg using stochastic gradient descent with gradients computed using back-prop\-agation~\cite{rumelhart1988learning}); (ii) validation: to tune the model's hyper-parameters, basically configuration parameters that can be modified to better fit the expected application workload; and 
(iii) evaluation: to evaluate the accuracy of the trained model \wrt to its predictions on other annotated data, to determine whether or not it predicts correctly.
Typically, training, validation, and testing data are drawn from the same initial dataset. 

For image classification, a DNN can be trained in either of the following two settings: 

\noindent
(i)~\textbf{Single-label Classification.}
In a traditional single-label classification problem, each datum is associated with a single label \textit{l} from a set of disjoint labels $L$ where $|L| > 1$. If $|L| = 2$, the classification problem is called a binary classification problem; if $|L| > 2$, it is a multi-class classification problem~\cite{tsoumakas2007multi}. Among some popular image classification datasets, MNIST, CIFAR-10/CIFAR-100~\cite{cifar} and ImageNet~\cite{ILSVRC15} are all single-label,
where each image can be categorized into only one class or outside that class. 

\noindent
(ii)~\textbf{Multi-label Classification.}
In a multi-label classification problem, each datum is associated with a set of labels \textit{Y} where $Y \subseteq L$. 
COCO\cite{lin2014microsoft} and imSitu\cite{yatskar2016} are popular datasets for multi-label classification. For example, an image from the \coco dataset can be labeled as \textit{car, person, traffic light and bicycle}. A multi-label classification model is supposed to predict all of \textit{car, person, traffic light and bicycle} from a single image that shows all of these kinds of objects.

Given any single- or multi-label classification task, DNN classifier software tries to learn the decision boundary between the classes\textemdash all members of a class, say $C_i$, should be categorized identically irrespective of their individual features, 
and members of another class, say $C_j$, should not be categorized to $C_i$~\cite{bengio2013representation}. The DNN represents the input image in an embedded space with the feature vector at a certain intermediate layer and uses the layers after as a classifier to classify these representations. The {\em class separation} between two classes estimates how well the DNN has learned to separate each class from the other.
If the embedded distance between two classes is too small compared to other classes, or lower than some pre-defined threshold, we assume that the DNN could not separate them from each other.

\section{Methodology}
\label{sec:method}

We give a detailed technical description of \tool. 
We describe a typical scenario where we envision our tool might be used in the following and design the methodology accordingly.

\begin{figure}
\centering
\includegraphics[width=1\columnwidth]{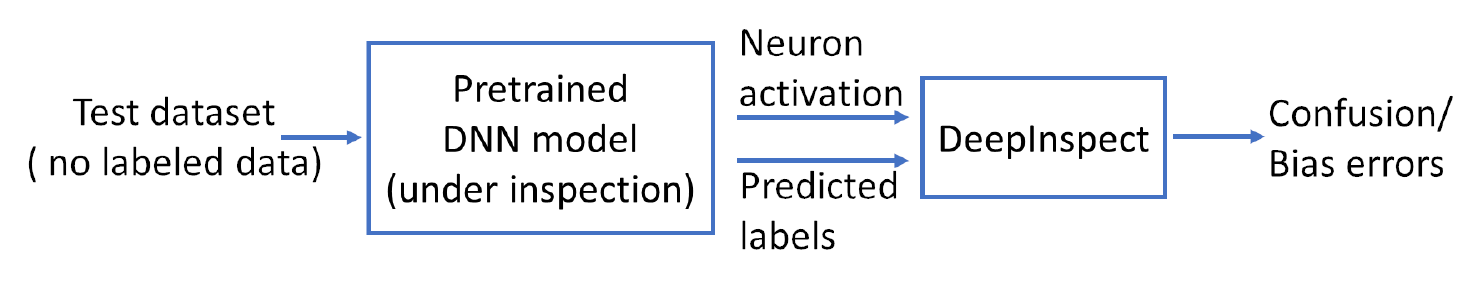}
\caption{\textbf{\small \tool Workflow}
}
\label{fig:workflow}
\end{figure}

\smallskip
\noindent
\textbf{Usage Scenario.} Similar to customer testing of post-release software, \tool works in a real-world 
setting where a customer gets a pre-trained model and tests its performance in a sample production scenario before deployment. The customer has white-box access to the model to profile, although all the data in the production system can be {\em unlabeled}. 
In the absence of ground truth labels, the classes are defined by the {\em predicted labels}. 
These predicted labels are used as class references as \tool tries to detect confusion and bias errors among the classes. 
\tool tracks the activated neurons per class and reports a potential class-level violation if the class-level activation-patterns are too similar between two classes. 
Such reported errors will help customers evaluate how much they can trust the model's results related to the affected classes. 
As elaborated in Section~\ref{sec:discussion}, once these errors are reported back to the developers, they can focus their debugging and fixing effort on these classes. 
Figure \ref{fig:workflow} shows the \tool workflow.

\subsection{Definitions}
 
Before we describe \tool's methodology in detail, we introduce definitions that we use in the rest of the paper.
The following table shows our notation. \\
\begin{center}
{\footnotesize
\begin{tabular}{ll}
    \toprule
    All neurons set     & $ N = \{N_1,N_2,...\}$ \\ 
    Activation function & $ out(N, c)$ returns output  \\ 
         & for neuron $N$ , input $c$. \\ 
    Activation threshold & $Th$ \\
    \bottomrule
\end{tabular}
}
\end{center}
\ \\
\noindent
    \textit{Neural-Path ($NP$).} 
For an input image $c$, we define {\em neural-path} as a sequence of neurons that are activated by $c$.

\noindent
\textit{Neural-Path per Class ($NP_{C}$).} 
For a class $C_i$, this metric represents a set consisting of the union of neural-paths activated by all the inputs in $C_i$. 


For example, consider a class {\tt cow} containing two images: a brown cow and a black cow. Let's assume they activate two neural-paths: $[N_1,N_2,N_3]$ and $[N_4,N_5,N_3]$. 
Thus, the neural-paths for class {\tt cow} would be 
$NP_{cow} = \{[N_1,N_2,N_3],[N_4,N_5,N_3]\}$. 
$NP_{cow}$ is further represented by a vector 
$(N_1^1,N_2^1,N_3^2,N_4^1,N_5^1)$, where the superscripts represent the number of times each neuron is activated by $C_{cow}$. 
Thus, each class $C_i$ in a dataset can be expressed with a {\em neuron activation frequency vector}, which captures how the model interacts with  $C_i$. 

\smallskip
\noindent
\textbf{\em Neuron Activation Probability}:  
Leveraging how {\em frequently} a neuron $N_j$ is activated by all the members from a class $C_i$, this metric estimates the probability of a neuron $N_j$ to be activated by $C_i$. Thus, 
we define: $\displaystyle P(N_j ~|~ C_i) = 
    \frac{|\{ c_{ik}~|~\forall{c_{ik}} \in C_i,out(N_j,c_{ik}) > Th\}|}{|C_i|}$

\medskip
\noindent
We then construct a $n\times m$ dimensional {\em neuron activation  probability matrix}, $\rho$, ($n$ is the number of neurons and $m$ is the number of classes) with its ij-th entry being $P(N_j ~|~ C_i)$.

{
\setlength{\abovedisplayskip}{0pt}
\setlength{\belowdisplayskip}{0pt}
\vspace{-0.3cm}
\begin{equation}
\scriptstyle
\label{eq:napm}
  \rho=
\begin{blockarray}{cccccc}
 & C_1 & ... & C_i & ... & C_m \\
\begin{block}{c(ccccc)}
  N_1 & p_{11} &   &   &   &  p_{1m} \\
  ... & ...    &   &   &   &  \\
  N_j & p_{j1} &   &...&   & p_{jm} \\
  ... & ...    &   &   &   &  \\
  N_n & p_{n1} &   &   &   &  p_{nm}\\
\end{block}
\end{blockarray}
\end{equation}
\vspace{-0.3cm}
}

This matrix  captures how a model interacts with a set of input data. 
The column vectors ($\rho_{\alpha m}$) represent 
 the interaction of a class $C_m$ with the model. 
Note that, in our setting, $C$s are predicted labels.

Since \napm ($\rho$) is designed to represent each class, it should be able to distinguish between different $C$s. 
Next, we use this metric to find two different classes of errors
often found in DNN systems: {\em confusion} and {\em bias} (see~\Cref{tab:example}).

\subsection{Finding Confusion Errors}
\label{sec:conf}

In an object classification task, when the model cannot distinguish one object class from another, confusion occurs. For example, as shown in~\Cref{tab:example}, a Google photo app model confuses a skier with the mountain.
Thus, finding confusion errors means checking how well the model can distinguish between objects of different classes. An error happens when the model under test classifies an object with a wrong class, or for multi-label classification task, predicts two classes but only one of them is present in the test image.

We argue that the model makes these errors because during the training process the model has not learned to distinguish well between the two classes, say $a$ and $b$. Therefore, the neurons activated by these objects are similar and the column vectors corresponding to these classes: $\rho_{\alpha a}$  and $\rho_{\alpha b}$ 
will be very close to each other. Thus, we compute the confusion score between two classes 
as the euclidean distance between their two probability vectors:  

{
\setlength{\abovedisplayskip}{0pt}
\setlength{\belowdisplayskip}{0pt}
\begin{equation}
\label{eq:conf}
\begin{split}
\scriptstyle
    \navd(a,b) = \Delta(a, b) = ||\rho_{\alpha a} - \rho_{\alpha b}||_2
\displaystyle
     = \sqrt{\sum_{i = 1}^{n} \left( P(N_i|a) - P(N_i|b)\right) ^{2}}
\end{split}
\end{equation}
}

If the $\Delta$ value is less than some pre-defined threshold (conf\_th) for two pairs of classes, the model will potentially make mistakes in distinguishing one from another, which results in confusion errors. 
This $\Delta$ is  called \navd  (\underline{N}euron \underline{A}ctivation \underline{P}robabiliy \underline{V}ector \underline{D}istance).

\subsection{Finding Bias Errors}
\label{sec:bias}

In an object classification task, bias occurs if the model under test shows disparate outcomes between two related classes. For example, 
we find that ResNet-34 pretrained by imSitu dataset, often mis-classifies a man with a baby as {\tt woman}. We observe that in the embedded matrix $\rho$, $\Delta(baby,woman)$ 
is much smaller than $\Delta(baby,man)$. 
Therefore, during testing, whenever the model finds an image with a baby, it is biased towards associating the baby image with a woman. Based on this observation, we propose an inter-class distance based metric to calculate the bias learned by the model. We define the bias between two classes $a$ and $b$  over a third class $c$ as follows: 
\begin{equation}
    \label{eq:bias}
  \displaystyle
    bias(a, b, c) := \frac{|\Delta(c, a) - \Delta(c, b)|}{\Delta(c, a) + \Delta(c, b)}
\end{equation}

 If a model treats objects of classes $a$ and $b$ similarly under the presence of a third object class $c$, $a$ and $b$ should have similar distance \wrt $c$ in the embedded space $\rho$; thus, the numerator of the above equation will be small. Intuitively, the model's output can be more influenced by the nearer object classes, \ie if $a$ and $b$ are closer to $c$. Thus, we normalize the disparity between the two distances to increase the influence of closer classes.
 
 This bias score is used to measure how differently the given model treats two classes in the presence of a third object class. 
 An \textbf{average bias} (abbreviated as \ab) between two objects $a$ and $b$ for all class objects $O$ is defined  as:
\begin{equation}
\label{eq:abgbias}
avg\_bias(a, b) := \frac{1}{|O|-2} \sum_{c\in O, c\neq a,b} bias (a, b, c)
\end{equation}
The above score captures the overall bias of the model between two classes. If the bias score is larger than some pre-defined threshold, we report potential {\em bias errors}.

Note that, even when the two classes $a$ and $b$ are not confused by the model, \ie $\Delta(a,b) > conf\_th$,  they can still show bias \wrt another class, say $c$, if  $\Delta(a,c)$ is very different from $\Delta(b,c)$. Thus, bias and confusion are two separate types of class-level errors that we intend to study in this work.

Using these above equations we develop a novel testing tool, \tool, to inspect a DNN implementing image classification tasks and look for potential confusion and bias errors. We implemented \tool in the Pytorch deep learning framework and Python 2.7. All our experiments were run on Ubuntu 18.04.2 with two TITAN Xp GPUs. For all of our experiments, we set the activation threshold $Th$ to be 0.5 for all datasets and models. We discuss why we choose 0.5 as neuron activation threshold and how different thresholds affect our performance in the section \ref{sec:discussion}.

\section{Experimental Design}
\label{sec:experiment}

\subsection{Study Subjects}
\label{sec:subj}

We apply \tool for both multi-label and single-label DNN-based classifications. Under different settings, \tool automatically inspects 8 DNN models for 6 datasets. 
Table~\ref{tab:subj} summarizes our study subjects. All the models we used are standard, widely-used models for each dataset.
We used pre-trained models as shown in the Table for all settings except for \coco with gender. For \coco with gender model, we used the gender labels from \cite{zhao2017men} and trained the model in the same way as \cite{zhao2017men}. imSitu model is a pre-trained ResNet-34 model \cite{yatskar2016}. There are in total 11,538 entities and 1,788 roles in the imSitu dataset. When inspecting a model trained using imSitu, we only considered the top 100 frequent entities or roles in the test dataset.

\input{tables/subject.tex}

Among the 8 DNN models, three  are pre-trained relatively more robust models that are trained using adversarial images along with regular images. These models are pre-trained by provably robust training approach proposed by~\cite{NIPS2018_8060}. Three models with different network structures are trained using the CIFAR10 dataset~\cite{NIPS2018_8060}.

\subsection{Constructing Ground Truth (GT) Errors}
\label{sec:gt}

To collect the ground truth for evaluating \tool, 
 we refer to the test images misclassified by a given model. We then aggregate these misclassified image instances by their real and predicted class-labels and estimate pair-wise confusion/bias. 

\subsubsection{GT of Confusion Errors}
\label{sec:gt_conf} 
Confusion occurs when a DNN often makes mistakes in disambiguating  
members of two different classes. In particular, if a DNN is confused between two classes, the classification error rate is higher between those two classes than between the rest of the class-pairs. Based on this, we define two types of confusion errors for single-label classification and multi-label classification separately:

\textit{\ta}: In single-label classification, Type1 confusion occurs when an object of class $x$ (\eg {\tt violin}) is misclassified to another class $y$ (\eg {\tt cello}).  
For all the objects of class $x$ and $y$, it can be quantified as: $\typeaconf(x, y) = \mean(\prob(x| y), \prob(y| x))$ \textemdash DNN's probability to misclassify class $y$ as $x$ and vice-versa, and takes the average value between the two. For example, given two classes {\tt cello} and {\tt violin}, $\typeaconf$ estimates the mean probability of {\tt violin} misclassified to {\tt cello} and vice versa. Note that, this is a bi-directional score, \ie misclassification of $y$ as $x$ is the same as misclassification of $x$ as $y$. 

\textit{\tb}: In multi-label classification, Type2 confusion occurs when an input image contains an object of class $x$ (\eg  {\tt mouse}) and no object of class $y$ (\eg {\tt keyboard}), but the model predicts both classes (see~\Cref{fig:coco_confusion_bugs}. For a pair of classes, this can be quantified as: $\typebconf(x, y) = \mean(\prob(( x, y)| x), \prob(( x, y)| y))$ to compute the probability to detect two objects in the presence of only one. For example, given two classes {\tt keyboard} and {\tt mouse}, $\typebconf$ estimates the mean probability of {\tt mouse} being predicted while predicting {\tt keyboard} and vice versa. 
This is also a bi-directional score. 
    
\begin{figure}
\centering
    \begin{subfigure}[b]{0.23\textwidth}
       \includegraphics[width=\linewidth]{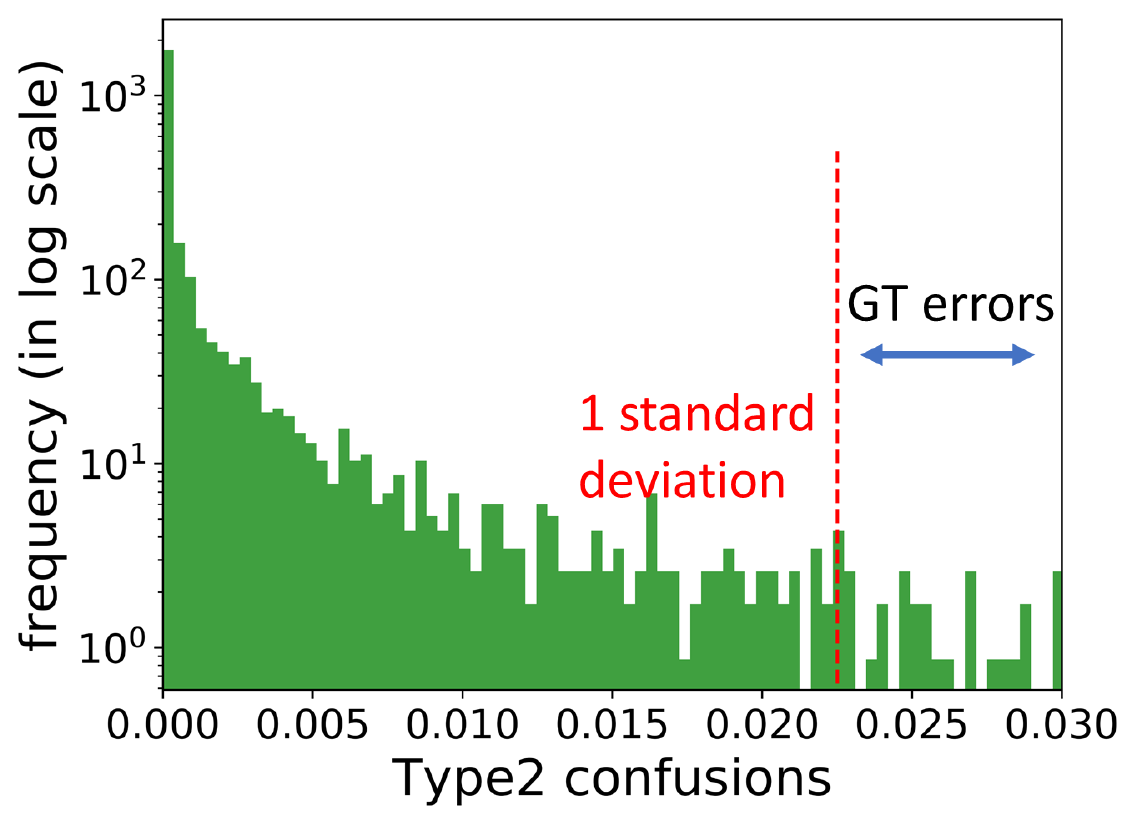}
        \caption{Confusions distribution}
        \label{fig:confusion_dist}
    \end{subfigure}
     ~ 
     \begin{subfigure}[b]{0.23\textwidth}
        \includegraphics[width=\linewidth]{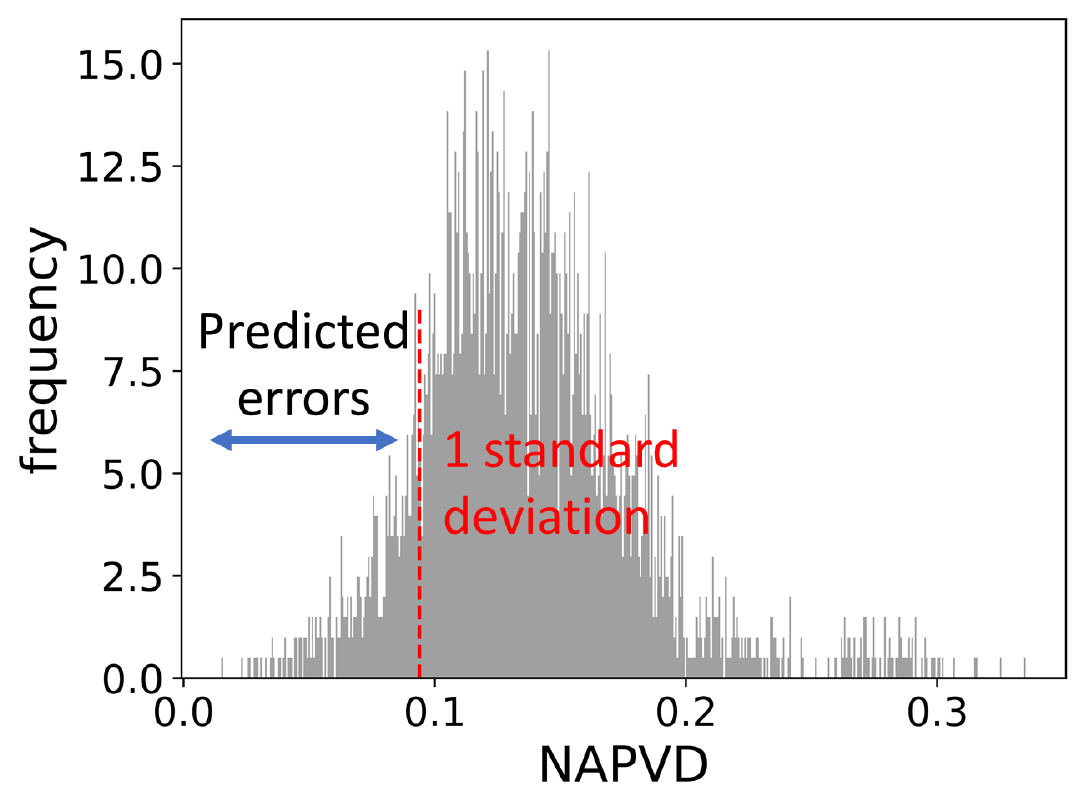} 
        \caption{NAPVD distribution}
        \label{fig:distance_dist}
    \end{subfigure}
\caption{\small{\textbf{Identifying \tb for multi-classification applications. LHS shows how we marked the ground truth errors based on Type2 confusion score. RHS shows \tool's predicted errors based on NAPVD score.}}
}
\label{fig:confusion_distribution}
\end{figure}

We measure $\typeaconf$ and $\typebconf$ by using a DNN's {\em true classification error} measured on a set of test images. 
They create the DNN's true confusion characteristics between all possible class-pairs. We then draw the distributions of $\typeaconf$ and $\typebconf$. For example, ~\Cref{fig:confusion_dist} shows $\typebconf$ distribution for \coco. The class-pairs with confusion scores greater than $1$ standard deviation from the mean-value are marked as pairs truly confused by the model and form our ground truth for confusion errors. For example, in the \coco dataset, there are 80 classes and thus 3160 class pairs (80*79/2); 178 class-pairs are ground-truth confusion errors.

Note that, unlike how a bug/error is defined in traditional software engineering, our suspicious confusion pairs have an inherent probabilistic nature. For example, even if $a$ and $b$ represent a confusion pair, it does not mean that all the images containing $a$ or $b$ will be misclassified by the model. Rather, it means that compared with other pairs, images containing $a$ or $b$ tend to have a higher chance to be misclassified by the model.

\subsubsection{GT of Bias Errors}
\label{sec:gt_bias} 
A DNN model is {\em biased} if it treats two classes differently.   
For example, consider three classes: {\tt man}, {\tt woman}, and {\tt surfboard}. An unbiased model should not have different error rates while classifying {\tt man} or {\tt woman} in the presence of {\tt surfboard}. 
To measure such bias formally, we define \textbf{confusion disparity} ($\cd$) to measure differences in error rate between classes $x$ and $z$ and between $y$ and $z$: $\cd(x, y, z) = |error(x, z) - error(y, z)|$, where the error measure can be either $\typeaconf$ or $\typebconf$ as defined earlier. $\cd$ essentially estimates the disparity of the model's error between classes $x$, $y$ (\eg~{\tt man}, {\tt woman}) \wrt a third class $z$ (\eg~{\tt surfboard}).

We also define an aggregated measure \textbf{average confusion disparity} (\textbf{avg\_cd}) between two classes $x$ and $y$ by summing up the bias between them over all third classes and taking the average:
\[
        \acd(x, y) := \frac{1}{|O|-2} \sum_{z \in O, z \neq x, y} \cd(x, y, z).
\]
Depending on the error types we used to estimate \acd, we refer to $Type1$\_\acd and $Type2$\_\acd. 
We measure \acd using the true classification error rate reported for the test images. Similar to confusion errors, we draw the distribution of \acd for all possible class pairs and then consider the pairs as {\em truly biased} if their \acd score is higher than one standard deviation from the mean value. Such truly biased pairs form our ground truth for bias errors. 

\subsection{Evaluating \tool}
\label{sec:eval_metric}

We evaluate \tool using a set of test images. 

\noindent
\textbf{Error Reporting.}
\tool reports confusion errors based on NAPVD (see~\Cref{eq:conf}) scores\textemdash lower NAPVD indicates errors.
We draw the distributions of NAPVDs for all possible class pairs, 
as shown in~\Cref{fig:distance_dist}. 
Class pairs having NAPVD scores lower than $1$ standard deviation from the mean score are marked as potential confusion errors. 

As discussed in~\Cref{sec:bias}, \tool reports bias errors based on \ab score (see~\Cref{eq:abgbias}), where higher \ab means class pairs are more prone to bias errors.
Similar to above, from the distribution of \ab scores, \tool predicts pairs with \ab greater than $1$ \textrm{standard deviation} from the mean score to be erroneous. 
Note that, while calculating error disparity between classes $a$, $b$ \wrt $c$ (see \Cref{eq:bias}), if both $a$ and $b$ are far from $c$ in the embedded space $\rho$, disparity of their distances ($\Delta$)  should not reflect true bias. Thus, while calculating $\ab(a,b)$ we further filter out the triplets where 
$\Delta (c,a)>th \land \Delta (c,b)>th$, where $th$ is some pre-defined threshold. In our experiment, we remove all the class-pairs having $\Delta$ larger than $1$ standard deviation (\ie $th$) from the mean value of all $Delta$s across all the class-pairs. 

\noindent
\textbf{Evaluation Metric.}\label{sec:eval_metric}
We evaluate \tool in two ways:

\noindent
\textbf{Precision \& Recall.}
We use precision and recall to measure \tool's accuracy. 
For each error type t, suppose that E is the number of errors detected
by \tool and A is the the number of true errors in the ground truth
set. Then the precision and recall of \tool are $\frac{|A\cap E|}{|E|}$ and $\frac{|A\cap E|}{|A|}$ respectively.

\noindent
\textbf{Area Under Cost Effective Curve (AUCEC).}
Similarly to how static analysis warnings are ranked based on their priority levels~\cite{rahman2013and}, we also rank the erroneous class-pairs identified by \tool based on the decreasing order of error proneness, \ie most error-prone pairs will be at the top. 
To evaluate the ranking we use a cost-effectiveness measure~\cite{Arisholm2010Systematic}, AUCEC (Area Under the Cost-Effectiveness Curve), which has become standard to evaluate rank-based bug-prediction systems~\cite{rahman2013sample,kamei2010revisiting,rahman2011bugcache, rahman2013and, ray2016naturalness}.

Cost-effectiveness evaluates when we inspect/test top n\% class-pairs in the ranked list (\ie inspection cost), how many true errors are found (\ie effectiveness). 
Both cost and effectiveness are normalized to 100\%.  
~\Cref{fig:confusion-ce} shows cost on the x-axis, and effectiveness on the y-axis,  indicating the portion of the ground truth errors found. AUCEC is the area under this curve. 

\noindent
\textbf{Baseline.} 
\label{sec:baseline}
We compare \tool \wrt two baselines:

\noindent
(i) MODE-inspired: A popular way to inspect each image is to inspect a  feature vector, which is an output of an intermediate layer ~\cite{Ma:2018:MAN:3236024.3236082,zhang2018deeproad}. 
However, abstracting a feature vector per image to the class level is non-trivial.
Instead, for a given layer, one could inspect the weight vector ($w_l = [w^0_l, w^1_l, ..., w^n_l]$) of a class, say $l$, where the superscripts represent a feature. Similar weight-vectors are used in MODE~\cite{Ma:2018:MAN:3236024.3236082} to compare the difference in feature importance between two image groups. 
In particular, from the last linear layer before the output layer we extract such per-class weight vectors and compute the pairwise distances between the weight vectors. Using these pairwise distances we calculate  confusion and bias metrics as described in Section~\ref{sec:method}. 
 
\noindent
(ii) Random: We also build a random model that picks random class-pairs for inspection~\cite{witten2005data} as a baseline. 

For AUCEC evaluation, we further show the performance of an optimal model that ranks the class-pairs perfectly\textemdash if $n$\% of all the class-pairs are truly erroneous, the optimal model would rank them at the top such that with lower inspection budget most of the errors will be detected. The optimal curve gives the lower upper bound of the ranking scheme. 

\noindent
\textbf{Research Questions.}
With this experimental setting, we investigate the following three research questions to evaluate \tool for DNN image classifiers: 

\begin{itemize}[leftmargin=*]
    \item 
    \textbf{RQ1.} 
    Can \tool distinguish between different classes?
    \item
    \textbf{RQ2.}~Can \tool identify the confusion errors?
    \item 
    \textbf{RQ3.}~Can \tool identify the bias errors?
\end{itemize}

\section{Results}
\label{sec:result}

We begin our investigation by checking whether de-facto neuron coverage-based metrics can capture class separation. 
 
\smallskip
\RQ{1}{\rqa}
\label{sec:rqa}

\noindent
\textbf{Motivation.} The heart of \tool's error detection technique lies in the fact that the underlying Neuron Activation Probability metric ($\rho$) captures each class abstraction reasonably well and thus  distinguishes between classes that do not suffer from class-level violations. In this RQ we check whether this is indeed true. We also check whether a new metric $\rho$ is necessary, \ie, whether existing neuron-coverage metrics could capture such class separations.

\noindent
\textbf{Approach.} We evaluate this RQ \wrt the training data since the DNN behaviors are not tainted with inaccuracies associated with the test images. Thus, all the class-pairs are benign. We evaluate this RQ in three settings: (i) using \tool's metrics, (ii) neuron-coverage proposed by  Pei~\etal~\cite{pei2017deepxplore}, and (iii) other neuron-activation related metrics proposed by DeepGauge~\cite{ma2018deepgauge}. 

\noindent
\textbf{Setting-1. \tool.}
Our metric, \napm ($\rho$), by construction is designed per class. 
Hence it would be unfair to directly measure its capability to distinguish between different classes. Thus, we pose this question in slightly a different way, as described below. 
For multi-label classification, each image contains multiple class-labels. 
For example, an image might have labels for both {\tt mouse} and {\tt keyboard}. 
Such coincidence of labels may create confusion\textemdash if two labels always appear together in the ground truth set, no classifier can distinguish between them. 
To check how many times two labels coincide, 
we define a coincidence score between two labels $L_a$ and $L_b$ as: 
$
\displaystyle
coincidence\left(L_a, L_b\right) = 
mean(P\left(L_a, L_b|L_a\right),P\left(L_a, L_b|L_b\right))
$.

The above formula computes the minimum probability of labels $L_a$ and $L_b$ occurring together in an image given that one of them is present. 
Note that this is a bi-directional score, \ie we treat the two labels similarly. 
The $mean$ operation ensures we detect the least coincidence in either direction. A low value of coincidence score indicates two class-labels are easy to separate and vice versa. 

Now, to check \tool's capability to capture class separation, we simply check the correlation between 
coincidence score and confusion score (\navd) 
from Equation~\ref{eq:conf} for all possible class-label pairs. 
Since only multi-label objects can have label coincidences, we perform this experiment for a pre-trained ResNet-50 model on the \coco multi-label classification task. 

A Spearman correlation coefficient between the confusion and coincidence scores reaches a value as high as 0.96, showing strong statistical significance. The result indicates that \tool can disambiguate most of the classes that have a low confusion scores. 




Interestingly, we found some pairs where coincidence score is high, but \tool was able to isolate them. For example, ({\tt cup,chair}), ({\tt toilet,sink}), \etc. Manually investigating such cases reveals that although these pairs often appear together in the input images, there are also enough instances when they appear by themselves. Thus, \tool disambiguates between these classes and puts them apart in the embedded space $\rho$. 
These results indicate \tool can also learn some hidden patterns from the context and, thus, 
can go beyond inspecting the training data coincidence for evaluating model bias/confusion, which is the de facto technique among machine learning researchers~\cite{zhao2017men}.

\begin{figure}[!phtb]
\centering
\includegraphics[width=0.8\columnwidth]{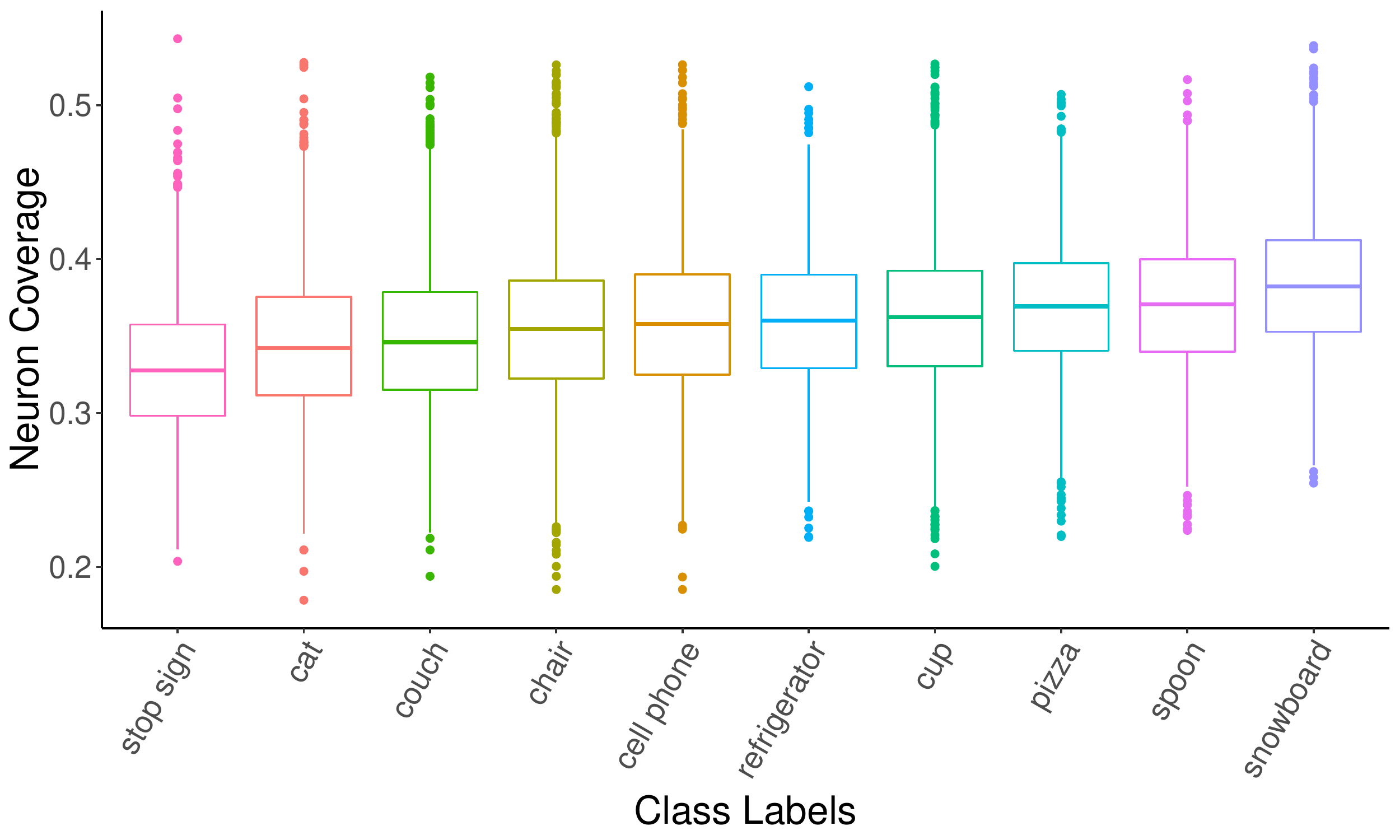}
\caption{\textbf{\small Distribution of neuron coverage per class label, for 10 randomly picked class labels, 
from the \coco dataset. 
}}
\label{fig:coverage}
\end{figure}

Next, we investigate whether popular white-box metrics can distinguish between different classes. 

\noindent
\textbf{Setting-2. Neuron Coverage ($NC$)}~\cite{pei2017deepxplore}
computes the ratio of the union of neurons activated by an input set and the total number of neurons in a DNN. 
Here we compute $NC$ per class-label, \ie for a given class-label, we measure the number of neurons activated by the images tagged with that label \wrt to the total neurons. The activation threshold we use is 0.5. 
We perform this experiment on \coco and \cifar to study multi- and single-label classifications. 
Figure~\ref{fig:coverage} shows results for \coco. We observe similar results for \cifar.

Each boxplot in the figure shows the distribution of neuron coverage per class-label across all the relevant images. 
These boxplots visually show that {\em different labels} have very {\em similar $NC$} distribution. 
We further compare these distributions using \mbox{Kruskal Test}~\cite{kruskaltest}, which is a non-parametric way of comparing more than two groups. Note that we choose a non-parametric measure as $NC$s may not follow normal distributions. (Kruskal Test is a parametric equivalent of the one-way analysis of variance (ANOVA).) The result reports a $p-value << 0.05$, \ie some differences exist across these distributions. 
However, a pairwise Cohend's effect size for each class-label pair, as shown in the following table, shows 
more than 56\% and 78\% class-pairs for \cifar and \coco have small to negligible effect size. 
This means neuron coverage cannot reliably distinguish a majority of the class-labels.

\begin{center}
{\scriptsize
    \begin{tabular}{l|rrrr}
    \toprule
         \multicolumn{5}{c}{  Effect Size of neuron coverage across different classes }  \\
    \toprule
    Exp Setting & negligible & {small} & {medium} & {large} \\
    \midrule
    \coco & 40.51\% & 38.19\% & 16.96\% & 4.34\% \\
    \cifar & 31.94\% & 25.69\% & 23.87\% & 18.48\% \\
    \bottomrule
    \end{tabular}%
}
\end{center}


\noindent
\textbf{Setting-3.~DeepGauge~\cite{ma2018deepgauge}.}
Ma \etal~\cite{ma2018deepgauge} argue that each neuron has a primary region of operation; they identify this region by using a boundary condition $[{low},{high}]$ on its output during training time; outputs outside this region ($(-\infty, {low}) \cup ({high}, +\infty)$) are marked as corner cases. They therefore introduce multi-granular neuron and layer-level coverage criteria. For neuron coverage they propose: (i) {\em k-multisection coverage} to evaluate how thoroughly the primary region of a neuron is covered, (ii) {\em boundary coverage} to compute how many corner cases are covered, and (iii) {\em strong neuron activation coverage} to measure how many corner case regions are covered in (${high}, +\infty$) region. For layer-level coverage, they define (iv) {\em top-k neuron coverage} to identify the most active k-neurons for each layer, and (v) {\em top-k neuron pattern} for each test-case to find a sequence of neurons from the top-k most active neurons across each layer.

We investigate whether each of these metrics can distinguish between different classes by measuring the above metrics for individual input classes following Ma \etal's methodology. 
We first profiled every neuron upper- and lower-bound for each class using the training images containing that class-label. 
Next, we computed per-class neuron coverage using test images containing that class; for k-multisection coverage we chose $k=100$ to scale up the analysis. It should be noted that we also tried $k=1000$ (which is used in the original DeepGauge paper) and observed similar results (not shown here).

\begin{figure}[!phtb]
\centering
\scriptsize
 \includegraphics[width=0.8\columnwidth]{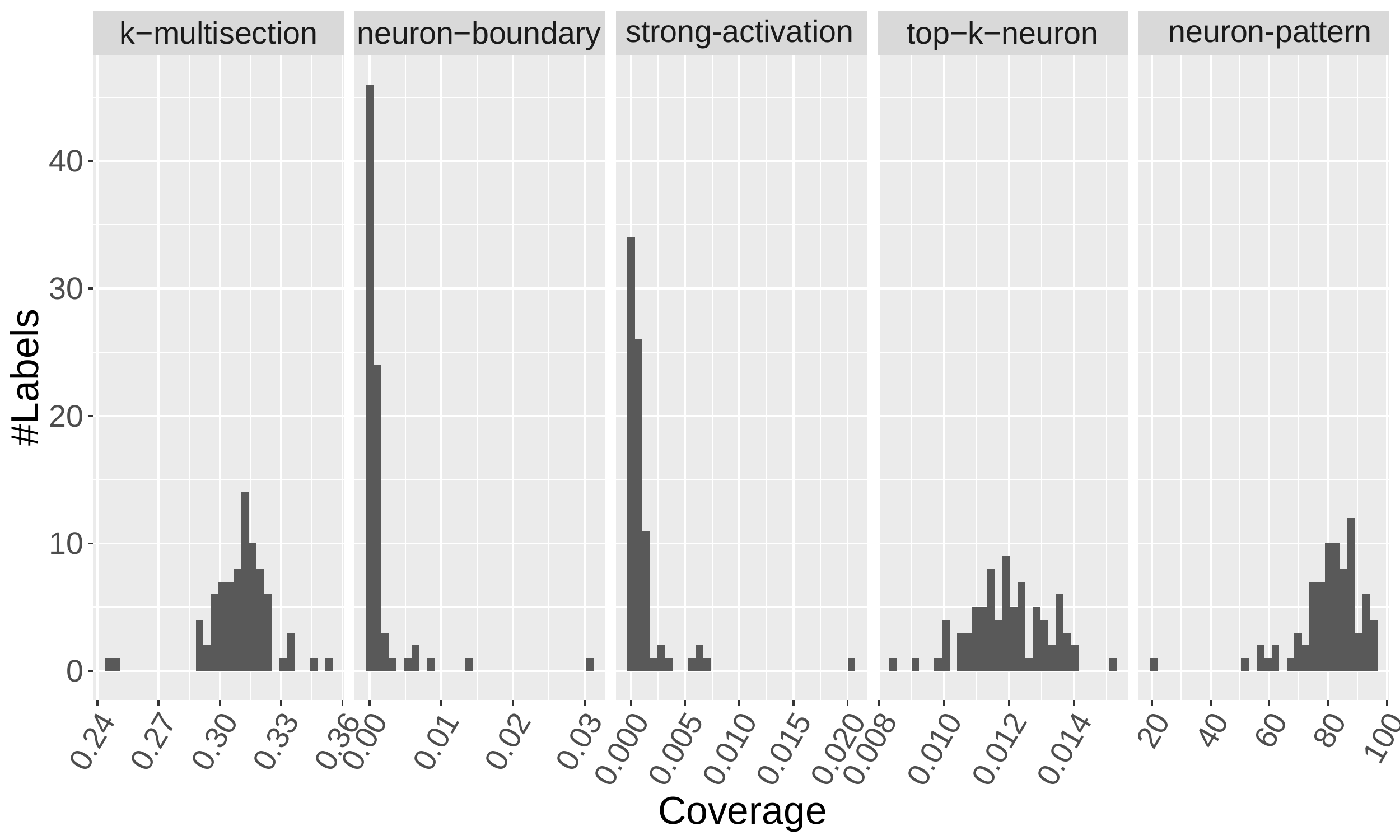}
\caption{\textbf{\small{Histogram of DeepGauge~\cite{ma2018deepgauge} multi-granular coverage per class label for \coco dataset}}
}
\label{fig:dg_coco}
\label{fig:dg}
\end{figure}

For layer-level coverage, we directly used the input images containing each class, where we select $k=1$.

\Cref{fig:dg_coco} shows the results as a histogram of the above five coverage criteria for the \coco dataset. For all five coverage criteria, there are many class-labels that share similar coverage. For example, in \coco, there are $52$ labels with k-multisection neuron coverage with values between $0.31$ and $0.32$. 
Similarly, there are $40$ labels with 0 neuron boundary coverage.
Therefore, none of the five coverage criteria are an effective way to distinguish between  different equivalence classes. The same conclusion was drawn for the \cifar dataset.

\RS{1}{\tool can disambiguate classes better than previous coverage-based metrics for the image classification task.}


We now investigate \tool's capability in detecting confusion and bias errors in DNN models. 
\input{tables/precision-recall.tex}
\RQ{2}{\rqb}
\label{sec:rq2}

\noindent
\textbf{Motivation.} To evaluate how well \tool can detect class-level violations, in this RQ, we report \tool's ability to detect the first type of violation, \ie, Type1/Type2 confusions \wrt to ground truth confusion errors, as described in~\Cref{sec:gt_conf}. 

\begin{figure}[h]
\centering
    \begin{subfigure}[b]{0.22\textwidth}
       \includegraphics[width=\linewidth]{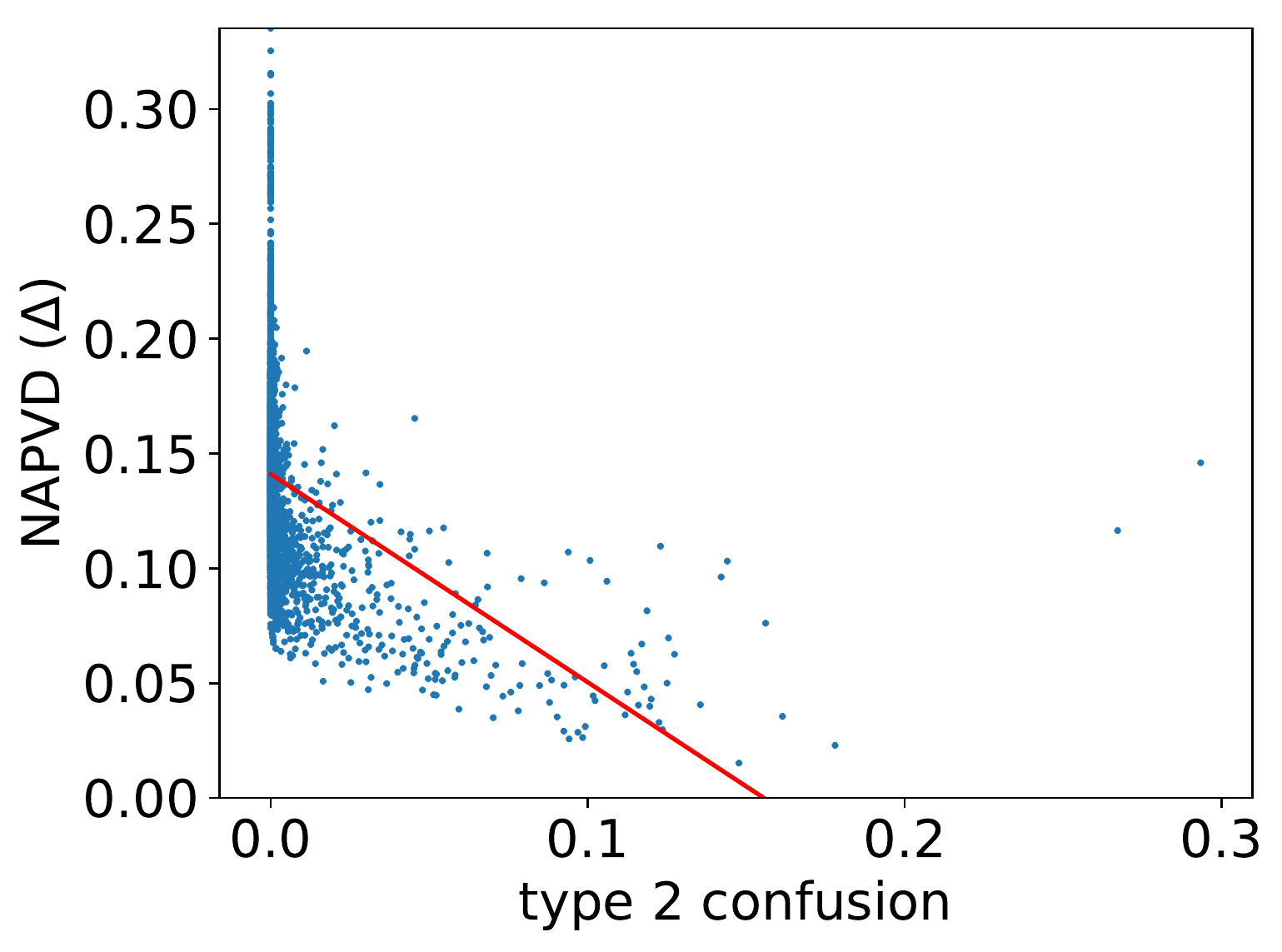}
        \caption{\coco dataset + ResNet-50}
        \label{fig:type2a}
    \end{subfigure}
    ~
    \begin{subfigure}[b]{0.22\textwidth}
       \includegraphics[width=\linewidth]{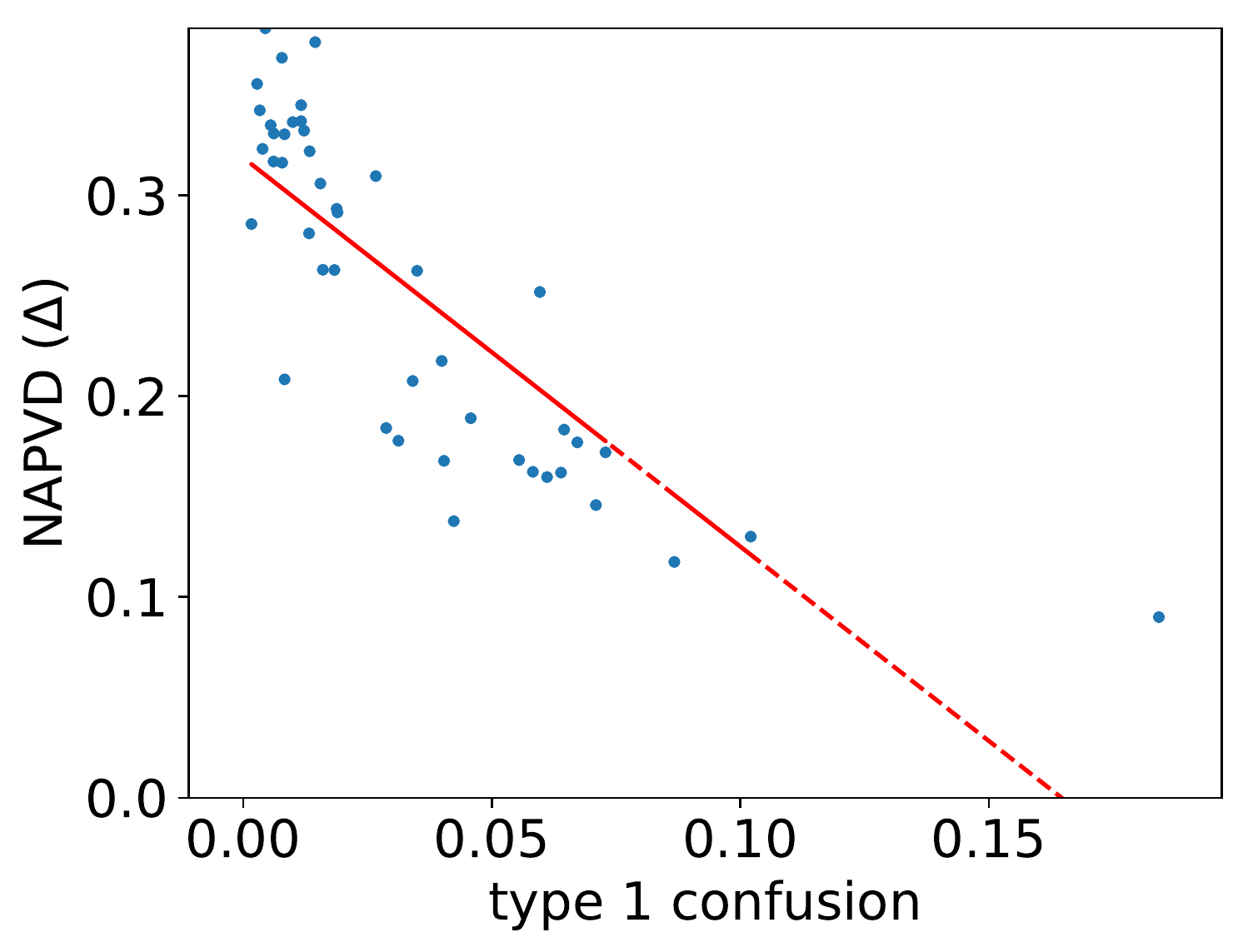}
        \caption{ Robust CIFAR-10 Small}
        \label{fig:type1a}
    \end{subfigure}
\caption{\textbf{\small{Strong negative Spearman correlation (-0.55 and -0.86) between   \navd and ground truth confusion scores.
}}}
\label{fig:rq2-corr}
\end{figure} 


We first explore the correlation between \navd and ground truth Type1/Type2 confusion score. Strong correlation has been found for all 8 experimental settings. ~\Cref{fig:rq2-corr} gives examples on COCO and CIFAR-10. These results indicate that \navd can be used to detect confusion errors\textemdash lower \navd means more confusion.


\input{tables/rq2}
\noindent
\textbf{Approach.} 
By default, \tool reports all the class-pairs with \navd scores one standard deviation less than the mean \navd score as error-prone (See~\Cref{fig:distance_dist}). In this setting, as the result shown on ~\Cref{tab:confusion_accuracy}, \tool reports errors at high recall under most settings. Specifically, on CIFAR-100 and robust CIFAR-10 ResNet, \tool can report errors as high as 71.8\%, and 100\%, respectively. 
\tool has identified thousands of confusion errors. 



\begin{figure}[!phtb]
\centering
\includegraphics[width=\linewidth]{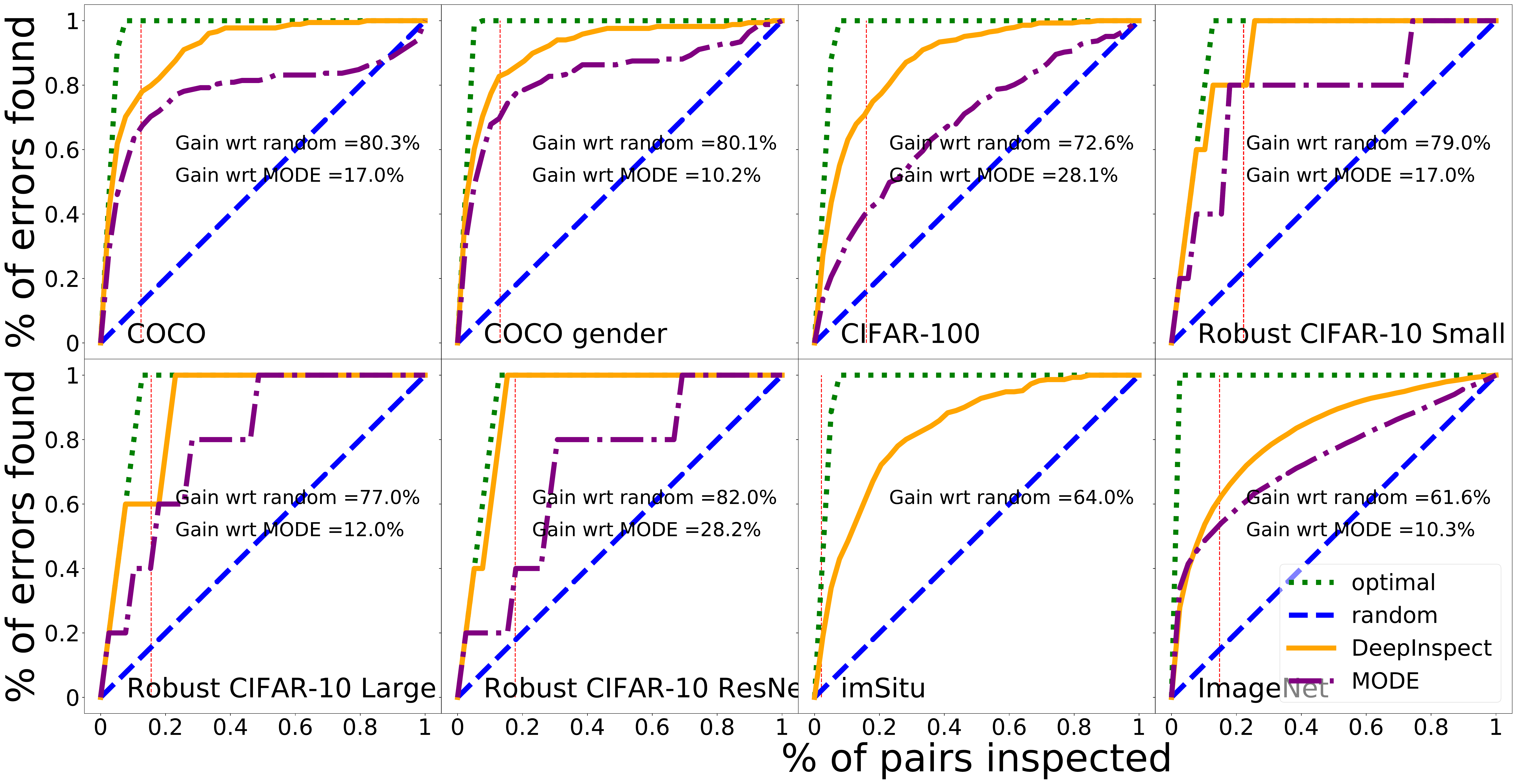}
\caption{\textbf{\small AUCEC plot of Type1/Type2 Confusion errors in three different settings. The \red{red} vertical line marks 1-standard deviation less from mean \navd score. \tool marks all class-pairs with \navd scores less than the red mark as potential errors.}}
\label{fig:confusion-ce}
\end{figure}

If higher precision is wanted, a user can choose to inspect only a small set of confused pairs based on \navd. As also shown in~\Cref{tab:confusion_accuracy}, when only the top1\% confusion errors are reported, a much higher precision is achieved for all the datasets. In particular, \tool identifies 31 and 39 confusion errors for the COCO model and the CIFAR-100 model with 100\% and 79.6\% precision, respectively. The trade-off between precision and recall can be found on the cost-effective curves shown on ~\Cref{fig:confusion-ce}, which show overall performance of \tool at different inspection cutoffs. Overall, \wrt a random baseline mode, \tool is gaining AUCEC performance from $61.6\%$ to $85.7\%$; \wrt a MODE baseline mode, \tool is gaining AUCEC performance from $10.2\%$ to $28.2\%$.

\begin{figure}
\centering

    \begin{subfigure}[b]{0.40\linewidth}
        \includegraphics[width=\linewidth]{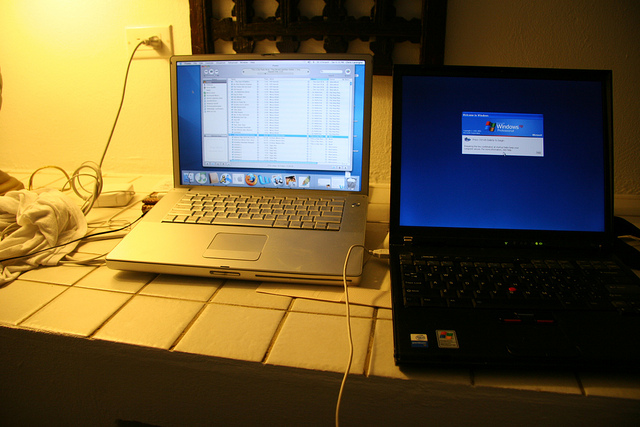}
        \caption{\small(keyboard,mouse)}
        \label{fig:confusionbugs1}
    \end{subfigure}
    ~
    \begin{subfigure}[b]{0.40\linewidth}
        \includegraphics[width=\linewidth]{{"figures/COCO_val2014_000000044294_oven-_microwave"}.jpg}
        \caption{\small (oven,microwave)}
        \label{fig:confusionbugs2}
    \end{subfigure}
    
\caption{\textbf{\small Confusion errors identified in \coco model. In each pair the second object is mistakenly identified by the model.}}
\label{fig:coco_confusion_bugs}
\vspace{-0.2cm}
\end{figure}

\begin{figure}[!htpb]
\centering
      \begin{subfigure}[b]{0.4\linewidth}
          \includegraphics[width=\linewidth]{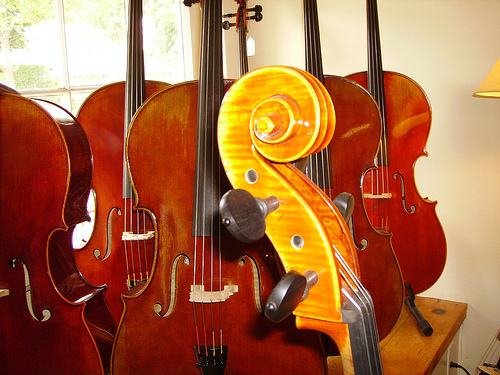}
          \caption{\small (cello, violin)}
          \label{fig:confusionbugs4}
      \end{subfigure}
      ~
    \begin{subfigure}[b]{0.4\linewidth}
        \includegraphics[width=\linewidth]{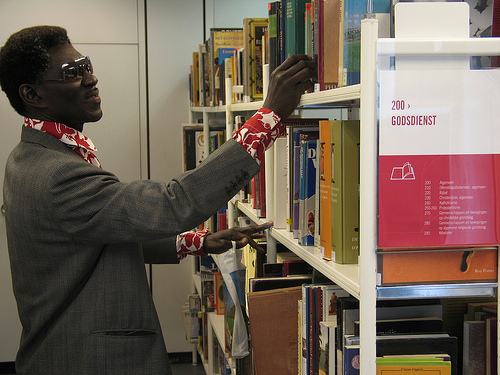}
        \caption{\small (library, bookshop)}
        \label{fig:confusionbugs5}
    \end{subfigure}
\caption{\textbf{\small Confusion errors identified in the ImageNet model. For each pair, the second object is mistakenly identified by the model.}
}
\label{fig:imagenet_confusion_bugs}
\end{figure}

Figure~\ref{fig:coco_confusion_bugs} and Figure~\ref{fig:imagenet_confusion_bugs} give some specific confusion errors found by \tool in the \coco and the ImageNet settings. In particular, as shown in Figure~\ref{fig:confusionbugs1}, when there is only a keyboard but no mouse in the image, the COCO model reports both. Similarly, Figure~\ref{fig:confusionbugs4} shows confusion errors on (cello, violin). There are several cellos in this image, but the model predicts it to show a violin.

Across all three relatively more robust 
CIFAR-10 models \tool identifies (cat, dog), (bird, deer) and (automobile, truck) as buggy pairs, where one class is very likely to be mistakenly classified as the other class of the pair. 
This indicates that these confusion errors are to be tied to the training data, so all the models trained on this dataset including the robust models may have these errors. These results further show that the confusion errors are orthogonal to the norm-based adversarial perturbations and we need a different technique to address them. 


We also note that the performance of all methods degrades quite a bit on ImageNet. ImageNet is known to have a complex structure, and all the tasks, including image classification and robust image classification \cite{Xie_2019_CVPR} usually have inferior performance compared with simpler datasets like CIFAR-10 or CIFAR-100. Due to such inherent complexity, the class representation in the embedded space is less accurate, and thus the relative distance between two classes may not correctly reflect a model’s confusion level between two classes.

\RS{2}{\tool can successfully find confusion errors with precision 21\% to 100\% at top1\% for both single- and multi-object classification tasks. 
\tool also finds confusion errors in robust models.}

\RQ{3}{\rqc}

\begin{figure}[!htpb]
\centering
\begin{subfigure}[b]{0.23\textwidth}
       \includegraphics[width=\linewidth]{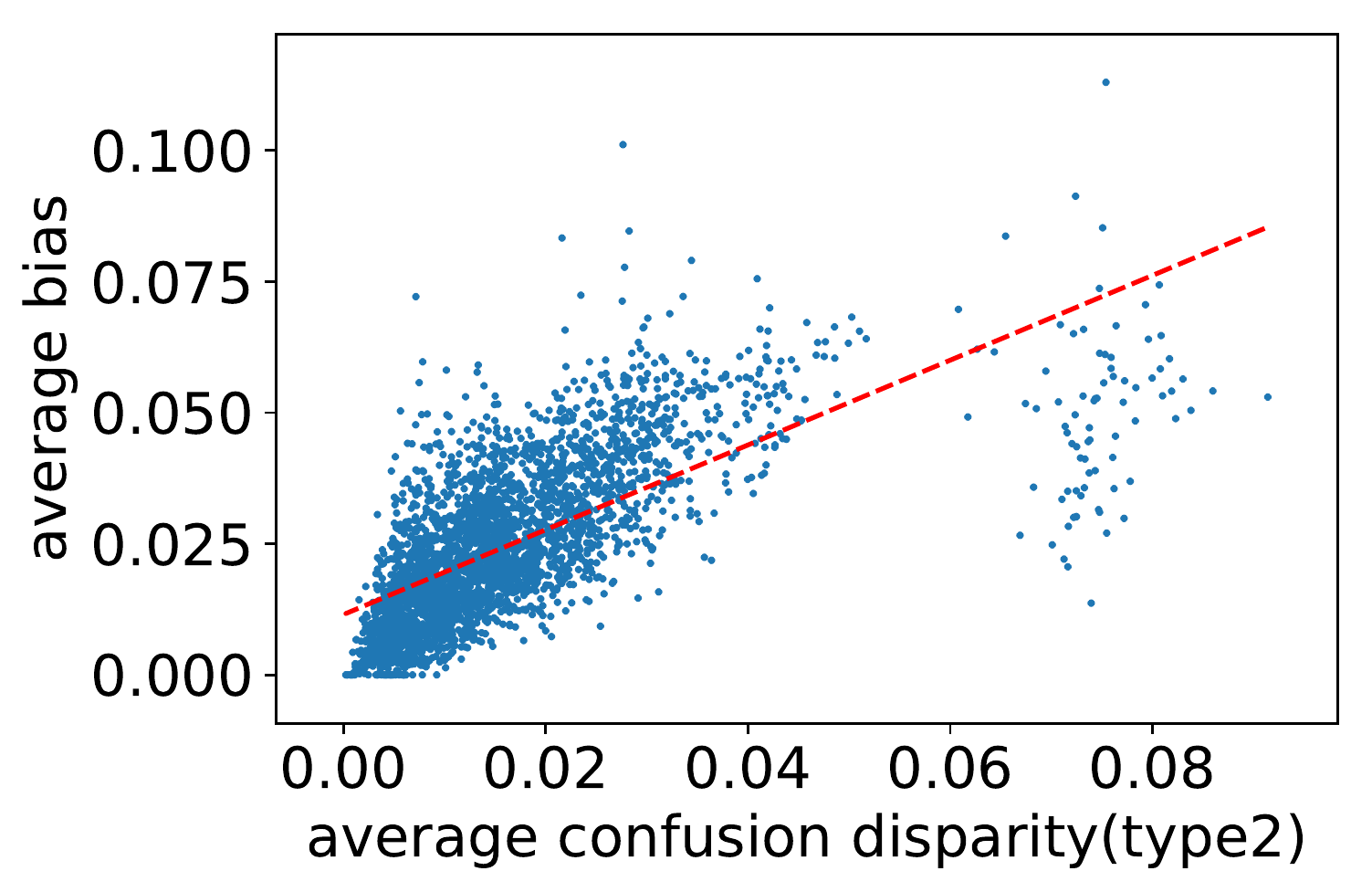}
        \caption{\coco}
        \label{fig:bias_coco_predicted_type2}
    \end{subfigure}
\begin{subfigure}[b]{0.23\textwidth}
        \includegraphics[width=\linewidth]{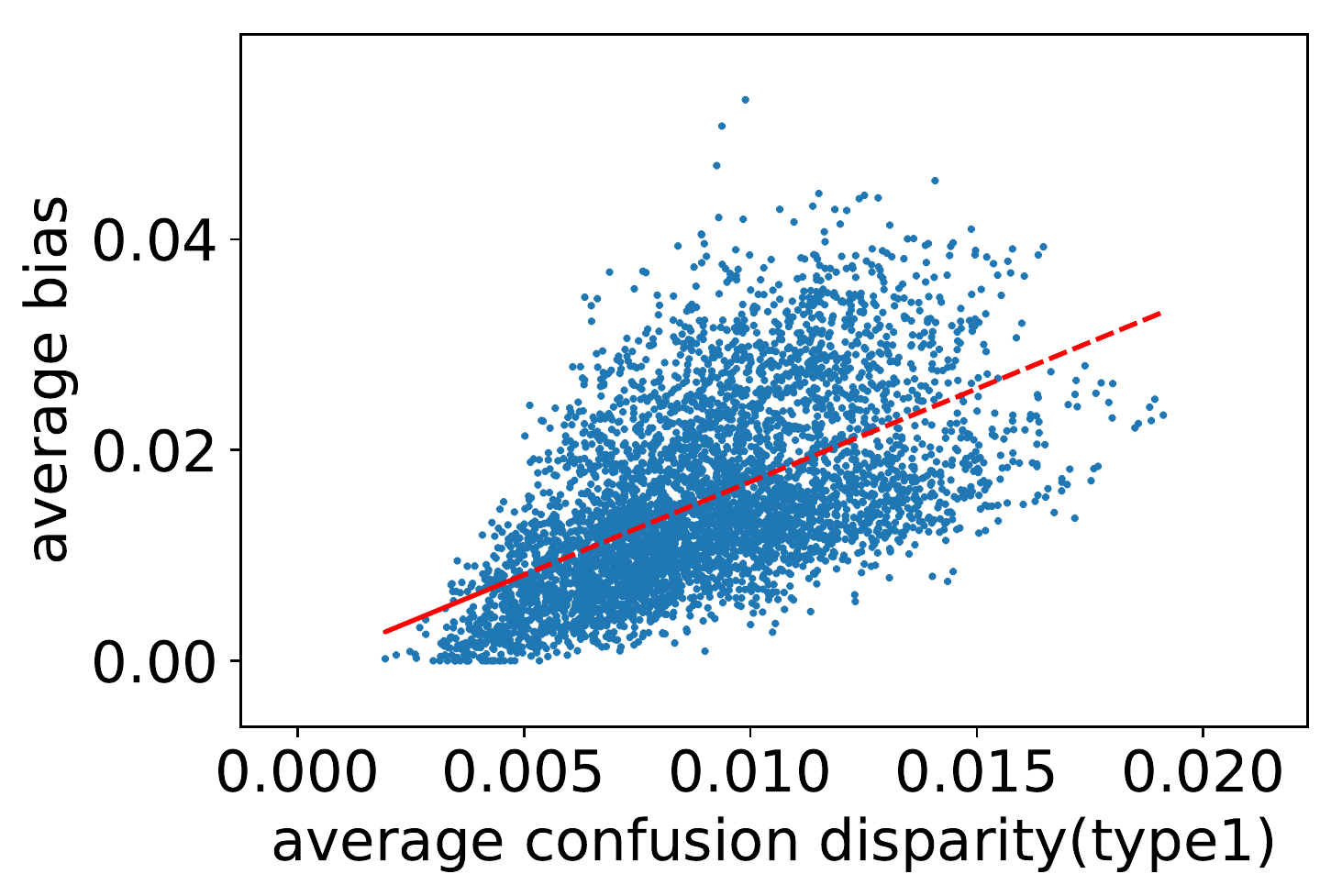}
        \caption{CIFAR-100}
        \label{fig:bias_coco_predicted_type1}
    \end{subfigure}
\caption{\textbf{\small Strong positive Spearman's correlation (0.76 and 0.62) exist between \acd and \ab while 
detecting classification bias.}
}
\label{fig:bias_coco_predicted}
\end{figure}

\noindent
\textbf{Motivation.} To assess \tool's ability to detect class-level violations, in this RQ, we report \tool's performance in detecting  the second type of violation, \ie, Bias errors as described in~\Cref{sec:gt_bias}.

\noindent
\textbf{Approach.} 
We evaluate this RQ by estimating a model's bias (\ab) using~\Cref{eq:abgbias} \wrt the ground truth (\acd), computed as in~\Cref{sec:gt_bias}. We first explore the correlation between pairwise \acd and our proposed pairwise \ab; 
~\Cref{fig:bias_coco_predicted} shows the results for \coco and CIFAR-10.
Similar trends 
were found in the other datasets we studied. 
The results show that a strong correlation exists between \acd and \ab. 
In other words, our proposed \ab is a good proxy for detecting confusion errors.

\input{tables/rq3}

\begin{figure}[h]
\centering
\includegraphics[width=\linewidth]{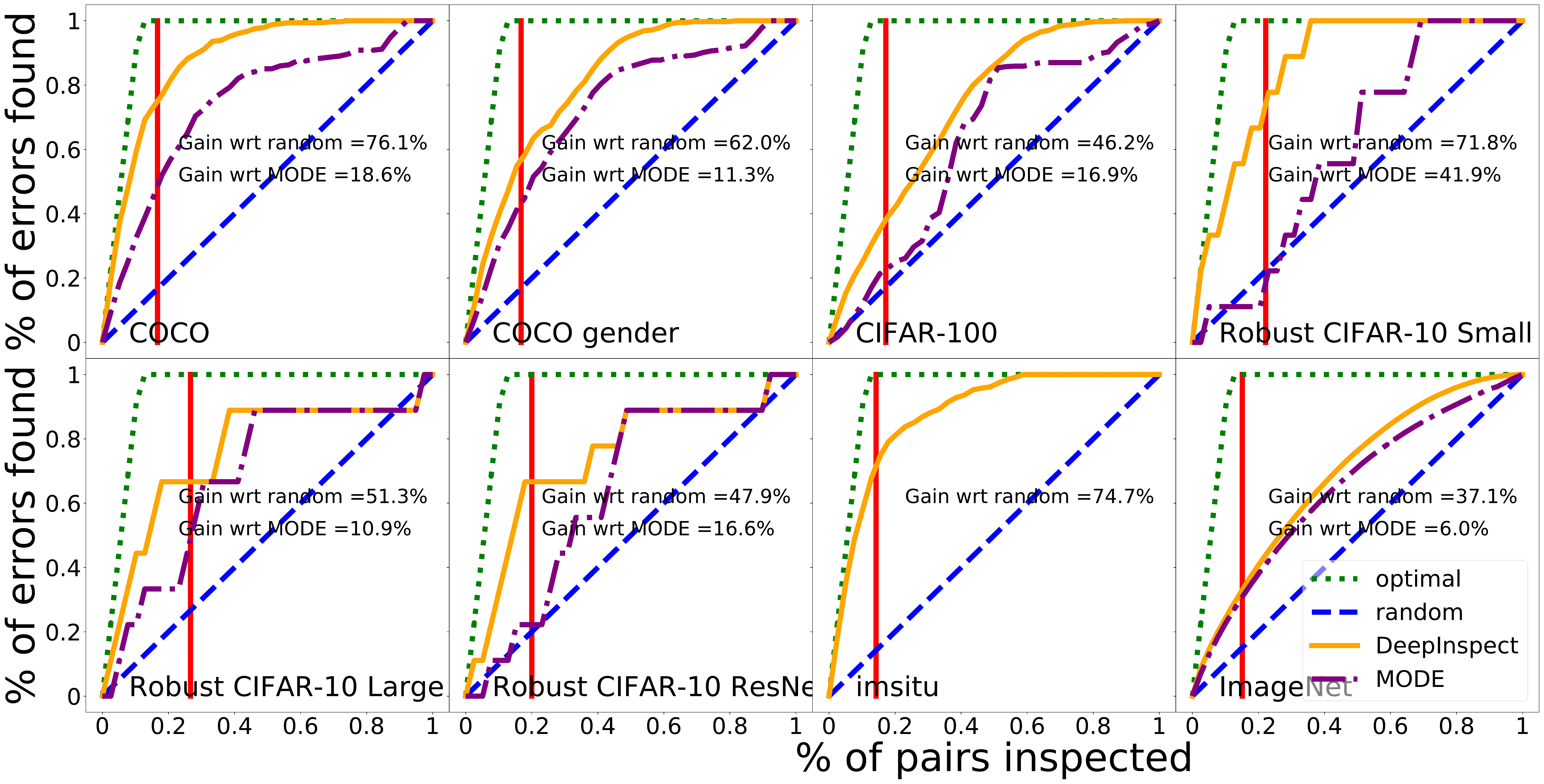}
\caption{\textbf{\small Bias errors detected \wrt the ground truth of \acd beyond one standard deviation from mean.}
}
\label{fig:bias-ce}
\end{figure}

As in RQ2, we also do a precision-recall analysis \wrt finding the bias errors across all the datasets.
We analyze the precision and recall of \tool when reporting bias errors at the cutoff Top1$\%$(\ab) and mean(\ab)+standard deviation(\ab), respectively. The results are shown in Table \ref{tab:bias_cutoffs}. At cutoff Top1$\%$(\ab), \tool 
detects suspicious pairs with precision as high as 75\% and 84\% for \coco and imSitu, respectively. At cutoff mean(\ab)+standard deviation(\ab), \tool has high recall but lower precision: \tool detects ground truth suspicious pairs with recall at 75.9\% and 71.8\% for \coco and imSitu.  \tool can report 657(=249+408) total true bias bugs across the two models. \tool outperforms the random baseline by a large margin at both cutoffs. As in the case of detecting confusion errors, there is a significant trade-off between precision and recall.
This can be customized based on user needs.
The cost-effectiveness analysis in ~\Cref{fig:bias-ce} shows the entire spectrum. 

As shown in Figure~\ref{fig:bias-ce}, \tool outperforms the baseline by a large margin. 
The AUCEC gains of \tool are from $37.1\%$ to $76.1\%$ w.r.t. the random baseline and from $6.0\%$ to $41.9\%$ w.r.t. the MODE baseline across the 8 settings. \tool's performance is close to the optimal curve under some settings, specifically the AUCEC gains of the optimal over \tool are only 7.11\% and 7.95\% under the \coco and ImSitu settings, respectively.

\begin{figure}[!htpb]
\centering
    \begin{subfigure}[b]{0.30\linewidth}
        \includegraphics[width=\linewidth]{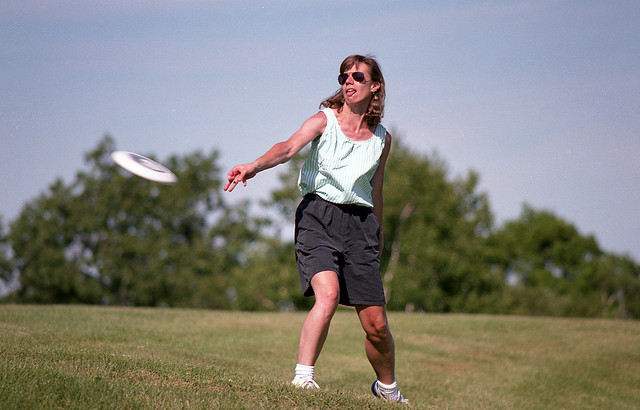}
    \end{subfigure}
    ~
    \begin{subfigure}[b]{0.30\linewidth}
        \includegraphics[width=\linewidth]{figures/bias_figures/COCO_val2014_000000493442}
    \end{subfigure}
    ~
    \begin{subfigure}[b]{0.30\linewidth}
       \includegraphics[width=\linewidth]{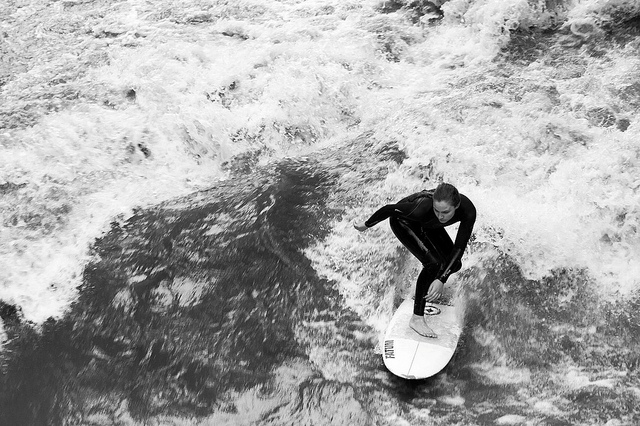}
    \end{subfigure}
\caption{\textbf{\small The model classifies the women in these pictures as men in the \coco dataset.}}
\label{fig:fairness_coco_predicted}
\end{figure}

\begin{figure}[!htpb]
\centering
    \begin{subfigure}[b]{0.25\linewidth}
        \includegraphics[width=\linewidth]{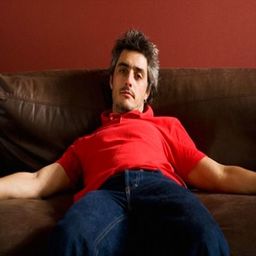}
    \end{subfigure}
    ~
    \begin{subfigure}[b]{0.25\linewidth}
        \includegraphics[width=\linewidth]{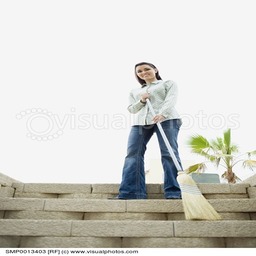}
    \end{subfigure}
\caption{\textbf{\small The model classifies the man in the first figure to be a woman and the woman in the second figure to be a man.}}
\label{fig:fairness_imsitu_predicted}
\end{figure}

Inspired by \cite{zhao2017men}, which shows bias exists between men and women in \coco for the gender image captioning task, we analyze the most biased third class $c$ for $a$ and $b$ being men and women. 
As shown in Figure \ref{fig:fairness_coco_predicted}, we found that sports like skiing, snowboarding, and surfboarding are more closely associated with men and thus misleads the model to predict the women in the images as men.  
Figure \ref{fig:fairness_imsitu_predicted} shows results on imSitu, where we found that the model tends to associate the class ``inside'' with women while associating the class ``outside'' with men.



We generalize the idea by choosing classes $a$ and $b$ to be any class-pair. We found that similar bias also exists in the single-label classification settings. For example, in ImageNet, one of the highest biases is between Eskimo\_dog and rapeseed \wrt Siberian\_husky. The model tends to confuse the two dogs but not Eskimo\_dog and rapeseed. This makes sense since Eskimo\_dog and Siberian\_husk are both dogs so more easily misclassified by the model.

One of the fairness violations of a DNN system can be drastic differences in accuracy across groups divided according to some sensitive feature(s). In black-box testing, the tester can get a number indicating the degree of fairness has been violated by feeding into the model a validation set. In contrast, \tool provides a new angle to the fairness violations. The neuron distance difference between two classes $a$ and $b$ \wrt a third class $c$ sheds light on why the model tends to be more likely to confuse between one of them and $c$ than the other. We leave a more comprehensive examination on interpreting bias/fairness violations for future work.

\RS{3}{\tool can successfully find bias errors for both single- and multi-label classification tasks, and even for the robust models, from 52\% to 84\% precision at top1\%.}

\section{Related Work}
\label{sec:rel}

\textbf{Software Testing \& Verification of DNNs.}
Prior research proposed different white-box testing criteria based on neuron coverage~\cite{pei2017deepxplore,ma2018deepgauge,tian2017deeptest} and neuron-pair coverage~\cite{sun2018testing}.
Sun \etal~\cite{sun2018concolic} presented a concolic testing approach for DNNs called DeepConcolic. They showed that their concolic testing approach can effectively increase coverage and find adversarial examples. Odena and Goodfellow proposed TensorFuzz\cite{pmlr-v97-odena19a}, which is a general tool that combines coverage-guided fuzzing with property-based testing to generate cases that violate a user-specified objective. It has applications like finding numerical errors in trained neural networks, exposing disagreements between neural networks and their quantized versions, surfacing broken loss functions in popular GitHub repositories, and making performance improvements to TensorFlow. 
There are also efforts to verify DNNs~\cite{pei2017towards,katz2017reluplex,huang2017safety,wang2018formal} against adversarial attacks. However, most of the verification efforts are limited to small DNNs and pixel-level properties. 
It is not obvious how to directly apply these techniques to detect class-level violations. 

\noindent
{\bf Adversarial Deep Learning.}
DNNs are known to be vulnerable to well-crafted inputs called adversarial examples, where the discrepancies are imperceptible to a human but can easily make DNNs fail~\cite{yuan2017adversarial,lu2017no, raghunathan2018certified, papernot2016cleverhans, evtimov2017robust, goodfellow2014explaining, kos2017adversarial, narodytska2016simple, nguyen2015deep, papernot2017practical, papernot2016limitations, szegedy2013intriguing, huang2017adversarial, kurakin2016adversarial}. 
Much work has been done to defend against adversarial attacks~\cite{bastani2016measuring, carlini2017towards, feinman2017detecting, grosse2017statistical, gu2014towards, metzen2017detecting, papernot2017extending, papernot2016distillation, shaham2015understanding, xu2017feature, zheng2016improving, he2017adversarial, metric_adv}.
Our methods have potential to identify adversarial inputs. 
Moreover, adversarial examples are usually out of distribution data and not realistic, while we can find both out-distribution and in-distribution corner cases. 
Further, we can identify a general weakness or bug rather than focusing on crafted attacks that often require a strong attacker model (\eg the attacker adds noise to a stop sign image).

\noindent
{\bf Interpreting DNNs.} 
There has been much research on model interpretability and visualization~\cite{lipton2016mythos, zhang2018visual, selvaraju2016grad,montavon2017methods,bau2017network,dong2017towards}. A comprehensive study is presented by Lipton ~\cite{lipton2016mythos}. 
Dong \etal ~\cite{dong2017towards} observed that instead of learning the semantic features of whole objects, neurons tend to react to different parts of the objects in a recurrent manner. Our probabilistic way of looking at neuron activation per class aims to capture holistic behavior of 
an entire class instead of an individual object so diverse features of class members can be captured. 
Closest to ours is by Papernot \etal~\cite{papernot2018deep}, who used nearest training points to explain adversarial attacks. 
In comparison, we analyze the DNN's dependencies on the entire training/testing data and represent it in \napm. We can explain the DNN's bias and weaknesses by inspecting this matrix.

\noindent
\textbf{Evaluating Models' Bias/Fairness.}
Evaluating the bias and fairness of a system is important both from a theoretical and a practical perspective~\cite{luong2011k, ZemelICML, zafar2017fairness, Brun:2018:SF:3236024.3264838}. 
Related studies first define a fairness criteria and then try to optimize the original objective while satisfying the fairness criteria ~\cite{Dwork:2011, Hardt2016, barocas-hardt-narayanan, DBLP:conf/fat/MenonW18, DBLP:conf/nips/DoniniOBSP18, DBLP:journals/corr/abs-1901-10837}. These properties are defined either at individual~\cite{Dwork:2011,Kusner,Kim} or group levels~\cite{Calders,Hardt2016,ZafarWWW}. 
In this work, we propose a definition of a bias error for image classification closely related to fairness notions at group-level. Class membership can be regarded as the sensitive feature and the equality that we want to achieve is for the confusion levels of two groups w.r.t. any third group. We showed the potential of \tool to detect such violations.

Galhotra \etal~\cite{galhotra2017fairness} first applied the notion of software testing to evaluating software fairness.  
They mutate the sensitive features of the inputs and check whether the output changes. One major problem with their proposed method, Themis, is that it assumes the model takes into account sensitive attribute(s) during training and inference. This assumption is not realistic since most existing fairness-aware models drop input-sensitive feature(s). Besides, Themis will not work on image classification, where the sensitive attribute (\eg, gender, race) is a visual concept that cannot be flipped easily. 
In our work, we use a white-box approach to measure the bias learned by the model during training. Our testing method does not require the model to take into account any sensitive feature(s). We propose a new fairness notion for the setting of multi-object classification, {\em average confusion disparity}, and a proxy, {\em average bias}, to measure for any deep learning model even when only unlabeled testing data is provided. In addition, our method tries to provide an explanation behind the discrimination. A complementary approach by Papernot \etal ~\cite{papernot2018deep} shows such explainability behind model bias in a single classification setting.

\section{Discussion \& Threats to Validity}
\label{sec:discussion}

\textbf{Discussion.}
In the 
literature, bug detection, debugging, and repair are usually three distinct tasks, and there is a large body of work investigating each 
separately.
In this work, we focus on bug detection for image classifier software.
A natural follow-up of our work will be debugging and repair leveraging \tool's bug detection. We present some preliminary results and thoughts.



A commonly used approach to improving (\ie fixing) image classifiers is active learning, which consists of  adding more labeled data by smartly choosing what to label next.
In our case, we can use \navd to identify the most confusing class pairs, 
and then target those pairs by collecting additional  examples that contain individual objects from the confusing pairs. 
We download 105 sample images from Google Images that contain isolated examples of these categories so that the model learns to disambiguate them.
We retrain the model from scratch using the original training data and these additional examples. Using this approach, we have some preliminary results on the \coco dataset. After retraining, we find that the $\typebconf$ of the top confused pairs reduces. For example, the $\typebconf$(baseball bat, baseball glove) is reduced from 0.23 to 0.16, and $\typebconf$(refrigerator, oven) is reduced from 0.14 to 0.10. Unlike traditional active learning approaches that encourage labeling additional examples near the current decision boundary of the classifier, our approach encourages the labeling of problematic examples based on confusion bugs. 

Another potential direction to explore is to use \tool in tandem with debugging \& repair tools for DNN models like MODE~\cite{Ma:2018:MAN:3236024.3236082}.
\tool enables the user to focus debugging effort on the vulnerable classes even in the absence of labeled data.  
For instance, once \tool identifies the vulnerable class-pairs, one can use the GAN-based approach proposed in MODE to generate more training data from these class-pairs, apply MODE to identify the most vulnerable features in these pairs to select for retraining.

We have also explored how the neuron coverage threshold($th$) used in computing $NAPVD$ affects our performance in detecting confusion and bias errors. We studied one multi-label classification task \coco~and one single-label classification task CIFAR-100. Table \ref{tab:confusion_coco_threshold}, \ref{tab:confusion_cifar100_threshold}, \ref{tab:bias_coco_threshold}, \ref{tab:bias_cifar100_threshold} show how our precision and recall change when using different neuron coverage thresholds ($th$). We observed that for CIFAR-100 and \coco~that DeepInspect’s accuracies are overall stable at $0.4\leq th \leq 0.75$. With smaller th($<0.25$), too many neurons are activated pulling the per-class activation-probability-vectors closer to each other. In contrast, with higher th($>0.75$), important activation information gets lost. Thus, we select $th=0.5$ for all the other experiments to avoid either issue.
\input{tables/threshold.tex}

\medskip
\noindent
\textbf{Threats to Validity.} We only test \tool on 6 datasets under 8 settings. We include both 
single-class and multi-class as well as regular and robust models to address these threats as much as possible.

Another limitation is that \tool needs to decide thresholds for both confusion errors and bias errors, and a threshold for discarding low-confusion triplets in the estimation of \ab. Instead of choosing fixed threshold, we mitigate this threat by choosing thresholds that are one standard deviation from the corresponding mean values and, also, reporting performance at top1\%. 




The task of accurately classifying any image 
is notoriously difficult. 
We simplify the problem by testing the DNN model only for the classes that it has seen during training. 
For example, while training, if a DNN does not learn to differentiate between black vs. brown {\tt cow}s (\ie, all the cow images only have label cow and they are treated as belonging to the same class by the DNN), \tool will not be able to test these sub-groups.

\section{Conclusion}
\label{sec:conclusion}
Our testing tool for DNN image classifiers, \tool, automatically detects confusion and bias errors in classification models. We applied \tool to six different popular image classification datasets and eight pretrained DNN models, including three so-called relatively more robust models.
We show that \tool can successfully detect class-level violations for both single- and multi-label classification models with high precision. 



%% file: tables/example.tex
\begin{table*}[!htbp]
\footnotesize
  \centering
  \caption{\textbf{\small Examples of real-world bugs reported in neural image classifiers}
  }
  \label{tab:example}
  \small
    \begin{tabular}{llll}
\textbf{Bug Type} & \textbf{Name}  & \textbf{Report Date} & \textbf{Outcome} \\
\toprule
  & Gorilla Tag~\cite{google_tag} & Jul 1, 2015 & Black people were tagged as gorillas by Google photo app. \\
\textbf{Confusion}  & Elephant is detected  & Aug 9, 2018 & Image Transplantation (replacing a sub-region of an image by   \\
  & in a room~\cite{elephant} &  & another image containing a trained object) leads to mis-classification. \\
  & Google Photo~\cite{google_photo} & Dec 10, 2018 & Google Photo confuses skier and mountain. \\
 \midrule
  & {Nikon Camera}~\cite{camera} & Jan 22, 2010 & Camera shows bias toward Caucasian faces when detecting people's blinks. \\
  & Men Like Shopping~\cite{zhao2017men} & July 29, 2017 & Multi-label object classification models show bias towards women on \\ 
\textbf{Bias} & & & activities like shopping, cooking, washing, \etc \\
  & {Gender Shades}\cite{Buolamwini2018GenderSI} & 2018  & Open-source face recognition services provided by IBM, Microsoft, and Face++  \\ 
  & & & have higher error rates on darker-skin females for gender classification. \\
\bottomrule
    \end{tabular}%
\end{table*}%

%% file: tables/subject.tex
\begin{table}[t]
  \setlength\tabcolsep{1pt} 
  \centering
  \scriptsize
  \caption{\textbf{\small Study Subjects}
  \vspace{-2mm}}
   \label{tab:subj}
  \begin{tabular}{l|l|r||r|r|r|r}
    \toprule
    \multicolumn{3}{c||}{\textbf{Dataset}} & \multicolumn{4}{c}{\textbf{Model}}\\ 
    \midrule
    \textbf{Classification} &  &  & \textbf{CNN} & &  & \textbf{Reported} \\
    \textbf{Task} & \textbf{Name} & \textbf{\#classes} & \textbf{Models} & \textbf{\#Neurons} & \textbf{\#Layers} & \textbf{Accuracy} \\
    \midrule
            & \coco\cite{lin2014microsoft} & 80 & ResNet-50\cite{zhao2017men} & 26,560 & 53 Conv &  0.73* \\
Multi-label & \coco gender\cite{zhao2017men} & 81 & ResNet-50\cite{he2016deep} & 26,560 & 53 Conv & 0.71* \\
  classification          & imSitu\cite{yatskar2016} & 205,095 & ResNet-34\cite{yatskar2016} & 8,448 & 36 Conv & 0.37\dag \\
  \midrule
      & CIFAR-100\cite{cifar} & 100  & CNN\cite{cifar100pretrained} & 2,916  & 26 & 0.74  \\ \cmidrule{2-7}
      & Robust  & 10 & Small CNN\cite{NIPS2018_8060} & 158 & 8 & 0.69  \\
Single-label   & CIFAR-10\cite{cifar}               &    & Large CNN\cite{NIPS2018_8060} & 1,226  & 14 & 0.73  \\
classification &                &    & ResNet\cite{NIPS2018_8060}    & 1,410  & 34 & 0.70  \\ \cmidrule{2-7}
      & \multirow{1}[0]{*}{ImageNet\cite{ILSVRC15}} & 1000  & ResNet-50\cite{simonyan2014very} & 26,560 & 53 Conv & 0.75  \\ 
    \bottomrule
    \end{tabular}%

{\small * reported in mean average precision, \dag reported in mean accuracy}
\end{table}


%% file: tables/rq2.tex

\begin{table}[htbp]
  \centering
  \scriptsize
  \setlength{\tabcolsep}{1pt}
  \caption{\small{\textbf{\tool performance on detecting confusion errors}}}
  \label{tab:confusion_accuracy}%
    \begin{tabular}{l|l|rrrr|rrrr}
    \toprule
          &       & \multicolumn{4}{c|}{\navd < mean-1std}        & \multicolumn{4}{c}{Top 1\%} \\
    \midrule
          &       & \textbf{TP}    & \textbf{FP}    & \textbf{Precision} & \textbf{Recall} & \textbf{TP}    & \textbf{FP}    & \textbf{Precision} & \textbf{Recall} \\
       \midrule
    COCO & \tool & 138   & 256   & 0.350  & 0.775 & 31    & 0     & 1     & 0.174 \\
         & MODE  & 126    & 382   & 0.248 & 0.708 & 26     & 5    & 0.839 & 0.146 \\
         & random & 22    & 372   & 0.056 & 0.124 & 1     & 30    & 0.032 & 0.006 \\
            \midrule
   COCO gender & \tool & 139   & 286   & 0.327 & 0.827 & 32    & 0     & 1     & 0.190 \\
        & MODE  & 125    & 379   & 0.248 & 0.744 & 30     & 2    & 0.938 & 0.179 \\
          & random & 22    & 403   & 0.052 & 0.131 & 1     & 31    & 0.031 & 0.006 \\
             \midrule
  CIFAR-100 & \tool & 206   & 584   & 0.261 & 0.718 & 39    & 10    & 0.796 & 0.136 \\
          & MODE & 111    & 605   & 0.155 & 0.387 & 22     & 27    & 0.449 & 0.077 \\
          & random & 45    & 745   & 0.057 & 0.157 & 2     & 47    & 0.041 & 0.007 \\
             \midrule
   R CIFAR-10 S & \tool & 4     & 6     & 0.400   & 0.800   & -     & -     & -     & - \\
          & MODE & 3     & 4     & 0.429   & 0.600   & -     & -     & -     & - \\
          & random & 1     & 9     & 0.100   & 0.200   & -     & -     & -     & - \\
             \midrule
 R CIFAR-10 L & \tool & 3     & 4     & 0.430  & 0.600   & -     & -     & -     & - \\
          & MODE & 3     & 5     & 0.375   & 0.600   & -     & -     & -     & - \\
          & random & 0     & 7     & 0     & 0     & -     & -     & -     & - \\
             \midrule
 R CIFAR-10 R & \tool & 5     & 3     & 0.625 & 1     & -     & -     & -     & - \\
          & MODE & 1     & 3     & 0.250   & 0.200   & -     & -     & -     & - \\
          & random & 0     & 8     & 0     & 0     & -     & -     & -     & - \\

             \midrule
   ImageNet & \tool & 4014  & 69957 & 0.054 & 0.617 & 1073  & 3922  & 0.215 & 0.165 \\
          & MODE & 3428   & 66987 & 0.049 & 0.527 & 1591    & 3404  & 0.319 & 0.245 \\
          & random & 962   & 73009 & 0.013 & 0.148 & 65    & 4930  & 0.013 & 0.010 \\
             \midrule
   imSitu & \tool & 48    & 58    & 0.453 & 0.165 & 31    & 19    & 0.620  & 0.107 \\
          & random & 6     & 100   & 0.057 & 0.020  & 2     & 48    & 0.040  & 0.007 \\
  \bottomrule
    \end{tabular}%
\end{table}%

%% file: tables/rq3.tex
\begin{table}[!htbp]
    \centering
  \scriptsize
      \caption{\textbf{\small{\tool performance on detecting bias errors}}} 
    \label{tab:bias_cutoffs}%
     \setlength{\tabcolsep}{1pt}
    \begin{tabular}{l|l|rrrr|rrrr}
  \toprule
          &       & \multicolumn{4}{c|}{\ab > mean+1std}        & \multicolumn{4}{c}{Top 1\%} \\
  \midrule
          &       & \textbf{TP} & \textbf{FP} & \textbf{Precision} & \textbf{Recall} & \textbf{TP} & \textbf{FP} & \textbf{Precision} & \textbf{Recall} \\
    \midrule
COCO & \tool & 249   & 278   & 0.472 & 0.759 & 24    & 8     & 0.75  & 0.073 \\
          & MODE  & 145    & 324   & 0.309 & 0.442 & 12     & 20    & 0.375 & 0.037 \\
          & random & 54    & 472   & 0.103 & 0.167 & 3     & 28    & 0.103 & 0.010 \\
    \midrule
COCO gender & \tool & 218   & 325   & 0.401 & 0.568 & 17    & 16    & 0.515 & 0.044 \\
          & MODE  & 151    & 328   & 0.315 & 0.393 & 13     & 20    & 0.394 & 0.034 \\
          & random & 64    & 478   & 0.118 & 0.168 & 3     & 28    & 0.118 & 0.010 \\
      \midrule
CIFAR-100 & \tool & 310   & 543   & 0.363 & 0.380  & 29    & 21    & 0.580  & 0.036 \\
          & MODE  & 69    & 315   & 0.180 & 0.085 & 5     & 45    & 0.100 & 0.001 \\
          & random & 140   & 711   & 0.165 & 0.172 & 8     & 41    & 0.165 & 0.010 \\
      \midrule
R CIFAR-10 S & \tool & 7     & 4     & 0.636 & 0.778 & -     & -     & -     & - \\
          & MODE  & 3    & 10   & 0.231 & 0.333 & -     & -    & - & - \\
          & random & 2     & 8     & 0.200   & 0.222 & -     & -     & -     & - \\
      \midrule
R CIFAR-10 L & \tool & 6     & 7     & 0.462 & 0.667 & -     & -     & -     & - \\
          & MODE   & 8    & 14   & 0.364 & 0.889 & -     & -    & - & - \\
          & random & 2     & 9     & 0.200   & 0.267 & -     & -     & -     & - \\
      \midrule
R CIFAR-10 R & \tool & 6     & 3     & 0.667 & 0.667 & -     & -     & -     & - \\
          & MODE   & 8    & 14   & 0.364 & 0.889 & -     & -    & - & - \\
          & random & 1     & 7     & 0.200   & 0.200   & -     & -     & -     & - \\
        \midrule

ImageNet & \tool & 26704 & 48913 & 0.353 & 0.330  & 3253  & 1742  & 0.651 & 0.040 \\
          & MODE   & 23881    & 47503   & 0.335 & 0.295 & 2355     & 2640    & 0.471 & 0.029 \\
          & random & 12234 & 63381 & 0.162 & 0.151 & 808   & 4186  & 0.162 & 0.010 \\
        \midrule
imSitu & \tool & 408   & 311   & 0.567 & 0.718 & 43    & 8     & 0.843 & 0.076 \\
          & random & 80    & 638   & 0.112 & 0.142 & 5     & 44    & 0.112 & 0.010 \\
      \bottomrule
    \end{tabular}%
\end{table}%

%% file: tables/threshold.tex
\begin{table}[htbp]
  \centering
  \scriptsize

  \setlength{\tabcolsep}{1pt}
  \caption{\small{\textbf{\tool impact of neuron coverage threshold on detecting confusion errors for COCO}}}
  \label{tab:confusion_coco_threshold}%
    \begin{tabular}{l|rrrr|rrrr}
    \toprule
NC threshold     & \multicolumn{4}{c|}{\navd < mean-1std}        & \multicolumn{4}{c}{Top 1\%} \\
    \midrule
  & \textbf{TP}    & \textbf{FP}    & \textbf{Precision} & \textbf{Recall} & \textbf{TP}    & \textbf{FP}    & \textbf{Precision} & \textbf{Recall} \\
       \midrule
 0.25 & 36   & 18   & 0.67  & 0.20 & 23    & 8     & 0.74     & 0.13 \\
         
            \midrule
 0.40 & 150   & 215   & 0.41 & 0.84 & 31    & 0     & 1     & 0.17 \\
       
             \midrule
 0.50 & 138   & 256   & 0.35 & 0.78 & 31    & 0    & 1 & 0.17 \\
         
             \midrule
0.60 & 137     & 264     & 0.34 & 0.77 & 30     & 1     & 0.97  & 0.17 \\
         
             \midrule
0.75 & 135     & 271     & 0.33  & 0.76   & 29     & 2   & 0.94 & 0.16 \\
  \bottomrule
    \end{tabular}%
\end{table}%

\begin{table}[htbp]
  \centering
  \scriptsize

  \setlength{\tabcolsep}{1pt}
  \caption{\small{\textbf{\tool impact of neuron coverage threshold on detecting confusion errors for CIFAR-100}}}
  \label{tab:confusion_cifar100_threshold}%
    \begin{tabular}{l|rrrr|rrrr}
    \toprule
NC threshold     & \multicolumn{4}{c|}{\navd < mean-1std}        & \multicolumn{4}{c}{Top 1\%} \\
    \midrule
  & \textbf{TP}    & \textbf{FP}    & \textbf{Precision} & \textbf{Recall} & \textbf{TP}    & \textbf{FP}    & \textbf{Precision} & \textbf{Recall} \\
       \midrule
 0.25 & 188   & 629   & 0.23  & 0.66 & 34    & 15     & 0.69     & 0.12 \\
         
            \midrule
 0.40 & 197   & 550   & 0.26 & 0.69 & 39    & 10     & 0.80     & 0.14 \\
       
             \midrule
 0.50 & 206   & 584   & 0.26 & 0.72 & 39    & 10     & 0.80     & 0.14 \\
         
             \midrule
0.60 & 211     & 596     & 0.26 & 0.74 & 37     & 12     & 0.76  & 0.13 \\
         
             \midrule
0.75 & 195     & 604     & 0.24  & 0.68   & 37     & 12     & 0.76  & 0.13 \\
  \bottomrule
    \end{tabular}%
\end{table}%

\begin{table}[htbp]
  \centering
  \scriptsize

  \setlength{\tabcolsep}{1pt}
  \caption{\small{\textbf{\tool impact of neuron coverage threshold on detecting bias errors for COCO}}}
  \label{tab:bias_coco_threshold}%
    \begin{tabular}{l|rrrr|rrrr}
    \toprule
NC threshold     & \multicolumn{4}{c|}{\ab > mean+1std}        & \multicolumn{4}{c}{Top 1\%} \\
    \midrule
  & \textbf{TP}    & \textbf{FP}    & \textbf{Precision} & \textbf{Recall} & \textbf{TP}    & \textbf{FP}    & \textbf{Precision} & \textbf{Recall} \\
       \midrule
 0.25 &  218 &  280  & 0.438  & 0.665 &  26   & 6     &  0.812    &0.079  \\
         
            \midrule
 0.40 &  260  & 275   &0.486  & 0.793 & 20    & 12     & 0.625     &0.061  \\
       
             \midrule
 0.50 &  249  & 278   &0.472  & 0.759 & 24    & 8    & 0.75 & 0.073 \\
         
             \midrule
0.60 &   190   &273  &0.410  & 0.579 &  24    & 8    & 0.75 & 0.073 \\
         
             \midrule
0.75 &  197    &54  & 0.785  & 0.601   &  32    & 0   & 1 & 0.098 \\
             \midrule
0.90 &  201    &102  & 0.663  & 0.592   &  32    & 0   & 1 & 0.094 \\
  \bottomrule
    \end{tabular}%
\end{table}%

\begin{table}[htbp]
  \centering
  \scriptsize

  \setlength{\tabcolsep}{1pt}
  \caption{\small{\textbf{\tool impact of neuron coverage threshold on detecting bias errors for CIFAR-100}}}
  \label{tab:bias_cifar100_threshold}%
    \begin{tabular}{l|rrrr|rrrr}
    \toprule
NC threshold     & \multicolumn{4}{c|}{\ab > mean+1std}        & \multicolumn{4}{c}{Top 1\%} \\
    \midrule
  & \textbf{TP}    & \textbf{FP}    & \textbf{Precision} & \textbf{Recall} & \textbf{TP}    & \textbf{FP}    & \textbf{Precision} & \textbf{Recall} \\
       \midrule
 0.25 & 289   & 569   & 0.337  & 0.355 & 18    & 32     & 0.36     & 0.022 \\
         
            \midrule
 0.40 & 272   & 545   & 0.333 & 0.334 & 27    & 23     & 0.54     & 0.033 \\
       
             \midrule
 0.50 & 310   & 543   & 0.363 & 0.380 & 29    & 21     & 0.58     & 0.036 \\
         
             \midrule
0.60 & 279     & 473     & 0.371 & 0.342 & 26     & 24     & 0.54  & 0.032 \\
         
             \midrule
0.75 & 276     & 455     & 0.378  & 0.339   & 29     & 21     & 0.58  & 0.036 \\
            \midrule
0.90 & 179     & 587     & 0.234  & 0.220   & 12     & 38     & 0.24  & 0.015 \\
  \bottomrule
    \end{tabular}%
\end{table}%